\documentclass[fleqn,usenatbib]{mnras}

\usepackage{newtxtext,newtxmath}

\usepackage[T1]{fontenc}

\newcommand{\unit}{UNITsims}
\newcommand{\hi}{\text{H\,\textsc{i}}}
\newcommand{\secref}[1]{\hyperref[#1]{Section~\ref*{#1}}}
\newcommand{\appref}[1]{\hyperref[#1]{Appendix~\ref*{#1}}}
\newcommand{\tabref}[1]{\hyperref[#1]{Table ~\ref*{#1}}}

\usepackage{graphicx}	
\usepackage{amsmath}	

\title[\hi\ IM anisotropic clustering and BAO]{
\begin{minipage}{7.03 in}
\vskip -0.4 in
\begin{flushright}
{\rm \small IFT-UAM/CSIC-21-63}
\end{flushright}
\end{minipage}\\
\vskip 0.0in
\vspace{0.1 cm} 
\hi\ intensity mapping correlation function from UNIT simulations: BAO and observationally induced anisotropy}
\author[S. Avila, B. Vos-Gin\'es et al.]{
Santiago Avila,$^{1, 2}$\thanks{santiago.avila@uam.es}
Bernhard Vos-Gin\'es,$^{1, 2}$\thanks{bernhard.vos@estudiante.uam.es}
Steven Cunnington,$^{3}$
Adam R.~H.~Stevens,$^{4}$
\newauthor
Gustavo Yepes,$^{1, 5}$
Alexander Knebe$^{1, 4, 5}$ 
and Chia-Hsun Chuang$^6$ \\
\\
$^{1}$ Departamento de F\'isica Te\'orica,  Universidad Aut\'onoma de Madrid, 28049 Madrid (Spain) \\
$^{2}$ Instituto de F\'isica Teorica UAM-CSIC, c/ Nicolás Cabrera 13-15, , 28049 Madrid \\
$^{3}$ School of Physics and Astronomy, Queen Mary University of London, Mile End Road, London E1 4NS, UK \\
$^{4}$ International Centre for Radio Astronomy Research, The University of Western Australia, Crawley, WA 6009, Australia\\
$^{5}$ Centro de Investigaci\'on Avanzada en F\'isica Fundamental (CIAFF), Facultad de Ciencias, Universidad Aut\'noma de Madrid, 28049 Madrid, Spain\\
$^{6}$ Kavli Institute for Particle Astrophysics and Cosmology, Stanford University, 452 Lomita Mall, Stanford, CA 94305, USA\\
}

\date{Accepted XXX. Received YYY; in original form ZZZ}

\pubyear{2021}

\begin{document}
\label{firstpage}
\pagerange{\pageref{firstpage}--\pageref{lastpage}}
\maketitle

\begin{abstract}
We study the clustering of \hi~intensity maps produced from simulations with a focus on baryonic acoustic oscillations (BAO) and the effects induced by telescope beam smoothing and foreground cleaning. We start by creating a \hi\ catalogue at $z=1.321$ based on the Semi-Analytic Galaxy Evolution (SAGE) model applied to the UNIT simulations. With this catalogue we investigate the relation between model \hi\ and the dark matter haloes and we also study the abundance of \hi, $\Omega_\hi$, predicted by this model. We then create synthetic \hi\ intensity maps with a Nearest-Grid-Point approach. In order to simulate the telescope beam effect, a Gaussian smoothing is applied on the plane perpendicular to the line of sight. The effect of foreground removal methods is simulated by exponentially damping the largest wavelength Fourier modes on the radial direction. We study the anisotropic 2-point correlation function (2PCF) $\xi(r_\perp,r_\parallel)$ and how it is affected by the aforementioned observational effects. In order to better isolate the BAO signal, we study several 2PCF $\mu$-wedges (with a restricted range of orientations $\mu$) tailored to address the systematics effects and we compare them with different definitions of radial 2PCFs. Finally, we discuss our findings in the context of an SKA-like survey, finding a clear BAO signal in most of the estimators here proposed. 
\end{abstract}




\section{Introduction}
\label{introduction}

$\Lambda$CDM is the current consensus model we have to describe the physical cosmology of our Universe. The model considers that only 4.9\% of its energy density is in form of ordinary matter. Dark matter (26.4\%) and dark energy (68.7\%) complete the rest of its energy density content \citep{planck2018} and determining their nature are still two of the greatest problems in modern cosmology. Dark energy was discovered when an unexpected acceleration of the Universe was revealed by Type-I supernova observations  \citep{riess,Perlmutter99} 
and confirmed by the Cosmic Microwave Background (CMB) \citep{bennett03} and Baryonic Acoustic Oscillations (BAO) \citep{eisenstein,cole2005}.  
Over the last two decades, larger and more precise campaigns have been run in order to better understand the cosmological model. Some of the current state-of-the-art cosmological observations are provided by the Planck satellite's measurements of the CMB \citep{planck2018},  PANTHEON's compilation of Type-Ia supernovae \citep{pantheon} 
and measurements of BAO from BOSS+eBOSS \citep{eboss}.
While these data allow us to place constraints below $1\%$ on some cosmological parameters\,---\,such as $\Omega_{\Lambda}$, $H_0$, and $\sigma_8$\,---\,the nature of both dark matter and dark energy is still a mystery. In this context, mapping a larger portion of the Large Scale Structure (LSS) of the Universe can provide more precise measurements. LSS can be particularly relevant to reveal information about the more recent Universe, whose expansion is dominated by dark energy.

In this work we focus on one of the most prominent and robust features in the LSS of the Universe: the Baryonic Acoustic Oscillations. 
BAO were generated in the early Universe, when the temperature was high enough to keep photons, electrons and baryons coupled, forming a plasma. While baryons are attracted towards dark-matter over-densities (already formed), photons apply pressure against them, generating waves that propagate until the recombination epoch. At recombination, the waves freeze due to the liberation of photons, leaving an imprint in the LSS. From that moment, the BAO feature will only grow due to the expansion of the Universe, described by the scale factor. As a consequence, BAO can be used as a cosmological standard ruler to constrain the angular-diameter distance $d_\text{A}(z)$ and the Hubble parameter $H(z)$, when observed in angular and radial scales, respectively.

The first BAO detection in the clustering of galaxies was achieved nearly simultaneously by the 2dF galaxy redshift survey \citep{percival01, cole2005} and the Sloan Digital Sky Survey (SDSS) \citep{eisenstein}. 
Later, more precise BAO measurements were achieved by surveys such as the 6dFGS \citep{Beutler11}, WiggleZ \citep{WiggleZ}, and the Baryonic Oscillation Spectroscopic Survey \citep[BOSS,][]{BOSS2012}, part of the SDSS series. The latter experiment, BOSS, meant a qualitative advance in the field of precision cosmology with LSS, reporting BAO diameter distance with an accuracy close to the $1\%$ \citep{BOSSfinal}. The extended BOSS (eBOSS) experiment was designed as part of the SDSS-IV program to explore BAO at higher redshifts, with its final results having been reported recently \citep{eboss}.

Whereas most of the aforementioned BAO measurements relied on spectroscopic redshift determinations to accurately map the positions of galaxies, there are other techniques to measure it. Both BOSS and eBOSS reported additional measurements of the BAO using the Lyman-$\alpha$ forest. This technique studies the distribution of gas densities along the line of sight of distant quasars by analysing the absorption lines that the former imprint in the spectra of the latter \citep{Bautista16,duMasduBourboux20}. 
The Dark Energy Survey (DES) also accurately measured the angular BAO by studying the clustering of galaxies using only photometric estimates of redshift \citep{DESY1BAO,DESY3BAO}. 
In this work, we study the viability of detecting BAO with the Square Kilometer Array using a promising technique: \hi\ intensity mapping (IM) \citep{Bharadwaj:2000av,Battye:2004re,Chang:2007xk,Wyithe:2007rq}.

The Square Kilometre Array (SKA) \citep{Bacon:2018dui} is a radio-telescope observatory under construction that will be able to map an unprecedented volume of the Universe by observing neutral hydrogen (\hi). 
The \hi\ emits photons with a wavelength of approximately 21\,cm when the parallel spins of the proton and electron (in the ground state) become antiparallel. For conducting large scale cosmology, the SKA will employ the IM technique, integrating all the \hi\ signal coming from an angular patch of the sky. However, when operating as an interferometer, the SKA dishes will not be tightly packed enough to provide a high sensitivity at the BAO scale \citep{Santos2015}. It therefore plans to operate in single-dish mode \citep{Battye:2012tg} where each of the 197 dishes operates as an auto-correlator i.e. a single telescope. This allows large volumes of the sky to be efficiently surveyed, whilst limiting thermal noise due to the integrated contribution from each single-dish. This process will result in a pixelised map (intensity map) with a relatively poor angular resolution due to the beam of the telescope, which is inversely proportional to the single-dish diameter. However, it provides an exquisite redshift resolution given by the frequency resolution used. 

Pathfinder surveys such as the Green Bank Telescope and Parkes have made \hi\ intensity mapping detections of cosmological signal (the LSS) in cross-correlation with optical surveys \citep{Masui:2012zc,Wolz:2015lwa,Anderson:2017ert, GBTeBOSS}. Furthermore, the MeerKAT radio telescope \citep{Santos:2017qgq}, a precursor and eventual component of the final SKA, has recently successfully calibrated the single-dish observations with an array of dishes in a pilot survey and is expected to produce \hi\ intensity maps soon \citep{Wang:2020lkn}. In the near future, the MeerKLASS survey is expected to run on that instrument, covering 4,000 deg$^2$ of the sky \citep{Santos:2017qgq}.

One can study the clustering of intensity maps to extract cosmological information from the LSS in a similar way to how we study the clustering of resolved galaxies. However, some differences are evident, and \citet{villaescusa} showed that a great part of the BAO signal is erased for a SKA-like \hi\ IM programs due to poor angular resolution. They showed that if one looks at the isotropic clustering, the BAO is completely erased. Instead, they proposed to study the 1D radial power spectrum of the intensity maps, which results in a visible BAO signature.

Another challenge in conducting precision cosmology with \hi\ IM is the presence of strong foregrounds, mostly coming from the Milky Way, which emits signals several orders of magnitude above the cosmological one. For conventional targeted galaxy surveys, these radio foregrounds are less of a contaminant but where the entire radiation from the sky is integrated, as in the IM technique, these foregrounds become dominant components in the observed signal. Several techniques have been proposed to remove the signal from the foregrounds  \citep[for an overview of those most commonly used, see][]{Cunnington:2020njn}, but the most robust ones tend to over-correct, removing any signal that is smooth with frequency. This efficiently removes the foregrounds, but also removes cosmological contributions from the largest scales along the radial direction. How these techniques modify the observed power spectrum of \hi\ and the inferred cosmological parameters has been studied in \citet{Wolz:2013wna,Shaw:2014khi,Alonso:2014dhk,Cunnington:2020wdu,soares}. 

In this work we analyse the anisotropic 2-point correlation function (2PCF) as measured on \hi\ Intensity Mapping simulations in the presence of both the telescope beam and the effects from foreground removal. Whereas past works have focused on either angular 2-point statistics or Cartesian Fourier space (mainly the power spectrum), we focus here on the Cartesian configuration space (the 2PCF). We additionally define a series of summary 2PCFs (as a function of a single distance) in order to isolate the BAO signal, giving particular attention to an SKA-like survey. Toward the final stages of this project, an independent team submitted a study based on theoretical models of the anisotropic 2PCF of \hi~IM \citep{Knennedy&Bull}. Whereas the focus is somewhat different and their configuration is focused on a MeerKLASS-like survey, some of their results are quite complementary to our findings in simulations, and simple visual comparison seem to indicate agreement between the two works. 

In this paper, we use the UNIT $N$-body simulations\footnote{\url{http://www.unitsims.org} \citep{Chuang}} (\unit), described in \secref{sec:unitsim}, which have been populated with galaxies with SAGE \citep{sage,Knebe21}, a semi-analytic model of galaxy formation. This procedure is further explained in \secref{sec:HIprescription}, where we also extract a \hi-to-dark-matter halo mass relation and compare it to previous studies. We also compare the total \hi\ predicted by these prescriptions $\Omega_\hi(z=1.321)$ with observational data in \secref{sec:omega}.
We then describe  how we create intensity maps out of the galaxy catalogues (\secref{sec:intensity_maps}), and we introduce two techniques to simulate the effect of the telescope beam (\secref{sec:beam}) and the effect of foreground-removal methods (\secref{sec:foregrounds}). In \secref{sec:2PCF} we describe the way we measure each of the isotropic [$\xi(r)$], the anisotropic [$\xi(r_\perp, r_ \parallel)$] and the multipole [$\xi_\ell (r)$] correlation functions from the simulations, and analyse observational effects on them.
In \secref{sec:BAO} we look at different ways to isolate the BAO signal in the presence of the aforementioned effects. We propose in \secref{sec:wedge} an orientation selection ($\mu$-wedged) in order to avoid the areas in $\{r_\perp, r_ \parallel\}$ space that are heavily affected by systematics. Different definitions of radial 2PCFs are probed in \secref{sec:radial}. Finally, we discuss how the different techniques proposed can isolate the BAO signal for an SKA-like survey in \secref{sec:nw}. Summary and conclusions are reported in \secref{sec:conclusions}.

\section{\hi\ Simulations}
\label{sec:simulation}
\subsection{The UNIT simulations}
\label{sec:unitsim}

Our work utilises the UNIT simulations (hereafter `\unit'), fully described in \cite{Chuang}. These are four $N$-body dark-matter-only simulations with cosmological parameters $\Omega_{\text{m},0}=1-\Omega_{\Lambda,0} = 0.3089$, $h = 0.6774$, $n_\text{s} = 0.9667$, and $\sigma_8 = 0.8147$. Each simulation has a comoving volume of $V_{\rm box} = 1.0\,h^{-3}\,{\rm Gpc}^3$, containing $4096^3$ particles of mass $1.25 \times 10^{9}\, h^{-1}\,{\rm M}_{\sun}$. 

The initial conditions were set at $z=99$ using the Zel\textquotesingle dovich Approximation \citep[][]{Zeldovich} with the public code \textsc{FastPM} \citep{fastpm} and particles were evolved to $z=0$ using the TreePM code \textsc{L-Gadget}, a version of \textsc{Gadget2} \citep[][]{springel} optimised for extreme memory parallelisation. 128 snapshots were output between the initial and final redshifts. In each of them, halo catalogues were created using the phase-space halo finder \textsc{rockstar} \citep{behroozi}.  
Subsequently, \textsc{ConsisitentTrees} \citep{consistenttrees} 
was run to obtain the merger trees that link the histories of all haloes across all snapshots.

One particularity of the \unit\ is that they are run with the \textit{fixed \& paired} technique \citep{Angulo2016}. This means that the initial conditions are not given by a random Gaussian field, but the modulus of the initial perturbations ($\lvert \delta_k\lvert$) is set to its expectation value determined by the initial power spectrum (\textit{fixed}). Additionally, simulations are run by pairs, with an offset of $\pi$ in their initial phases (\textit{paired}). Both of these techniques contribute to having a reduced scatter in the final average 2-point statistics. This allows us to obtain precise measurements of the 2PCF, even when using a relatively small box. \citet{Chuang} estimates the effective volume of this suite of simulations as $V_{\rm eff} \simeq 150\, h^{-3}\, {\rm Gpc}^3$,  which is about one third of the volume that the SKA \hi~IM program is expected to survey. Several studies have shown that, apart from reducing the covariance of the 2-point statistics (hence, the 4-point statistics), this technique does not introduce any bias on the 1-point statistics (halo mass function) the 2-point statistics  ($P(k)$ or $\xi(r)$) or  the $3$-point statistics \citep{Angulo2016,Villaescusa2018,Chuang}.
For this study, we work with the snapshot at redshift $z=1.321$ and use all boxes available: UNITSIM1, UNITSIM1\_InvPhase, UNITSIM2 and UNITSIM2\_InvPhase, always showing the 2PCF averaged over all (twelve, when we consider three different rotations for each box, see \autoref{sec:2PCF}) realisations. When shown, error bars are computed as the standard deviation of all available 2PCFs. We remark that these errorbars would not be representative of any particular experiment due to the reduction in the scatter mentioned above. Nevertheless, the relative size of different summary 2PCF will be very insightful, as we will see in \secref{sec:BAO}. \footnote{An alternative error bar would correspond to the standard deviation of the average of each pair. This yields a smaller and noisier error bar. However, we have checked that by using this definition we would reach qualitatively the same conclusions. We remark again that we will focus of the relative comparison on the error bars among different summary statistics.}

\subsection{From dark matter to \hi}
\label{sec:HIprescription}

Recently, the Semi-Analytic Galaxy Evolution model \citep[SAGE;][]{sage} was run on the UNITsims, as described in \citet{Knebe21}.
This code takes the merger history and mass accretion history of (sub)haloes to build the baryonic properties of the galaxies hosted by them. 
The model includes (but is not limited to) processes such as gas cooling and accretion, star formation with chemical enrichment and stellar feedback, and black-hole growth with active galactic nucleus (AGN) feedback.
 These physical processes establish a series of coupled differential equations that describe how baryonic mass shifts between discrete reservoirs (e.g.~a hot gas halo, a cold gas disc, a stellar disc, a stellar bulge, etcetera).  Galaxies are therefore modelled macroscopically in terms of integrated properties.  While the total mass in each reservoir of each galaxy is evolved numerically, the distribution of mass within each reservoir is described analytically.

SAGE nominally has eight free parameters.  These are manually calibrated to primarily reproduce the $z\!=\!0$ stellar mass function. Further, secondary constraints that were monitored during calibration include the $z\!=\!0$ black hole--bulge mass relation, the $z\!=\!0$ stellar mass--gas metallicity relation, the $z\!=\!0$ baryonic Tully--Fisher relation, and the cosmic star formation history.
The specific parameter set for the model used here is the same as that calibrated for the {\sc MultiDark-Galaxies} project \citep{knebe18}.  Only two parameters were tweaked from the original model of \citet{sage}.
We refer the reader to \citet{sage,knebe18}, and references therein for further details.

Following \citet{Knebe21}, we only consider objects that pass the following stellar mass cut: $M_{\star} \geq 10^{9}\,h^{-1}\,{\rm M}_{\sun}$. We also remove spurious objects with cold gas mass below $M_{\rm CG} = 10^2\,h^{-1}\,{\rm M}_{\sun}$ and we consider a minimum halo mass of $M_\text{h} \geq 10^{10.5}\,h^{-1}\,{\rm M}_{\sun}$ (a rounded number corresponding to $\sim26$ dark matter particles, slightly above the limit of 20 particles often used). 

 While SAGE does model the cold gas in galaxies, it does not specifically model the amount of that gas that is in the form of \hi.  We therefore must calculate this in post-processing.
Assuming that all the cold gas mass $M_{\rm CG}$ provided by SAGE is neutral, we can estimate the \hi\ mass $M_\hi$ by considering the mass fraction of Hydrogen over all possible elements $f_\text{H}=0.75$ (from Big Bang nucleosynthesis, BBN) and the molecular-to-atomic ratio $R_\text{mol} = M_{\text{H}_2}/M_\hi$:
\begin{equation}
\label{eq:HIprescription}
    M_\hi = f_\text{H}\cdot  \left(1-\frac{R_{\rm mol}}{R_{\rm mol}+1}\right) \cdot M_{\rm CG} \, \, .
\end{equation}

Following the discussion in, for example, \citet{zoldan}, we now consider two models for $R_{\rm mol}$: 
\begin{enumerate}
    \item A constant fraction of \hi, given by
    \begin{equation}
    \label{eq:Rmol04}
    R_{\rm mol}=0.4 \, \, .
    \end{equation}
    This relation dates back to local measurements performed by \citet{Zwaan2005} and \citet{Keres2003}, which was incorporated in the \hi\ modelling in \citet{Baugh04} and \citet{Power10}. This is the assumption made in previous IM studies \citep{cunnington20,Cunnington:2021czb} and will be our default option here (see discussion at the end of the subsection).

    \item A fitting function for \hi\ fraction given by \citet[][see their equations 10 \& 15]{Obreschkow2009}, based on \citet{blitz}, where they study and model both the \hi\ and H$_2$ gas profiles along the disks of galaxies. That prescription yields a $R_{\rm mol}$ dependent on the disk size ($r_{\rm disk}$), cold gas mass and disk stellar mass ($M_{\star , {\rm disk}}$) of each galaxy:  
    \begin{equation}
    \begin{split}
        \label{eq:Blitz}
        R_{\rm mol}= \Big( 3.44 R_c^{-0.506} + 4.82 R_c^{-1.054}\Big)^{-1} \, \, , \\
        {\rm with} \,  R_c = \big[K\ r_{\rm disk} ^{\ -4}\ M_{\rm CG}\ (M_{\rm CG} + 0.4 M_{\star , {\rm disk}}) \big]^{0.8}\, 
    \end{split}    
    \end{equation}
    representing the $R_{\rm mol}$ at the centre of the galaxy and $K=11.3\ {\rm m}^4\ {\rm kg}^{-2}$ being a constant.
    
\end{enumerate}

\begin{figure}
    \centering
    \includegraphics[trim=80 0 44 20, clip, width=0.48    \textwidth]{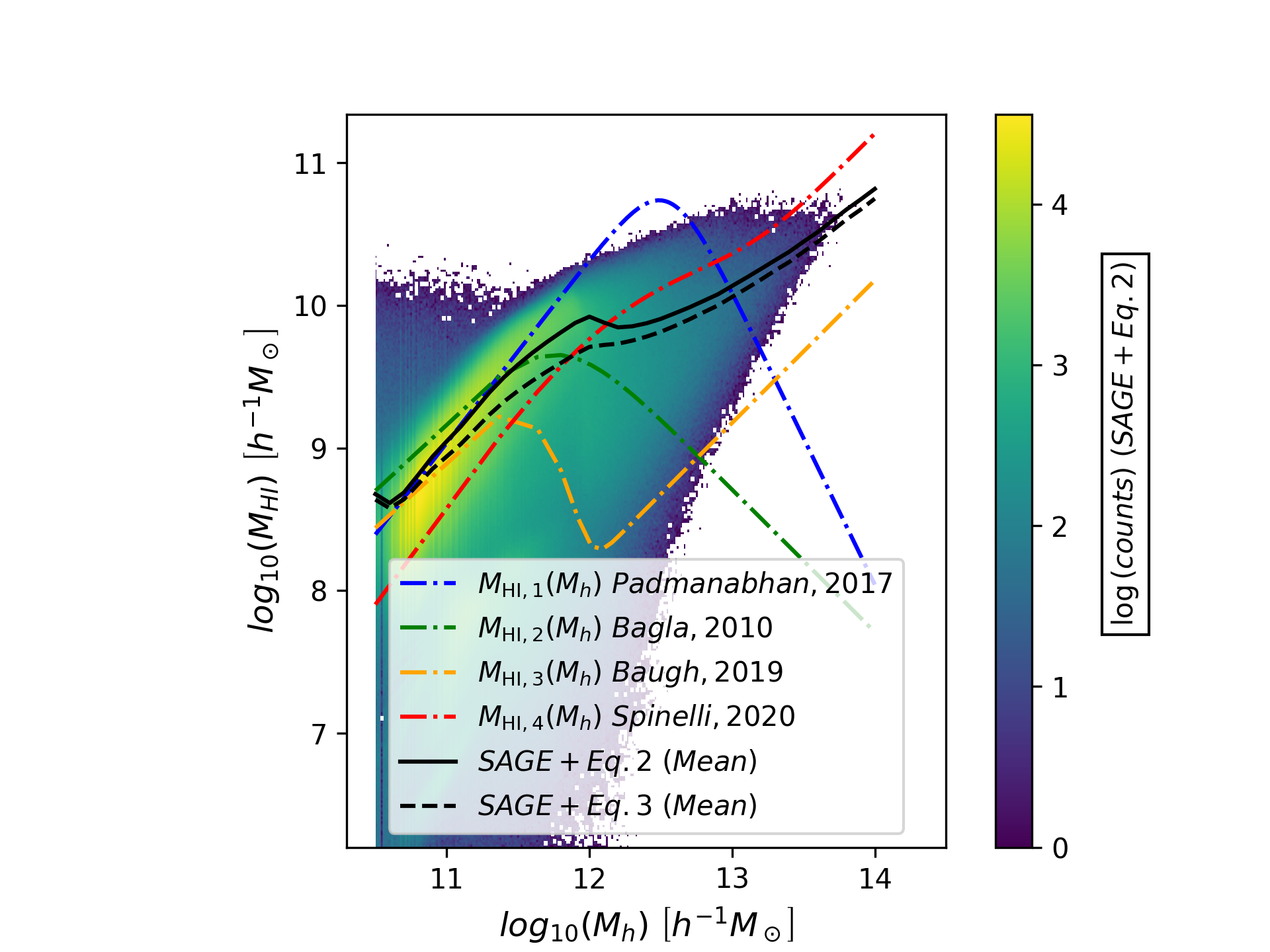}
    \caption{Colour map representing a 2D histogram of the main halo masses ($x$-axis) and total (summing all galaxies within a halo) \hi\ mass ($y$-axis, using the $R_{\rm mol}=0.4$ prescription) at $z=1.321$ for the \unit-SAGE catalogues. The solid black line represents the mean \hi\ mass for all galaxies within a given halo mass bin for $R_{\rm mol}=0.4$, whereas the dashed black line represents the same mean for the \citet{Obreschkow2009} prescription. We compare our findings to previous \hi--halo relations studied in the literature, as indicated in the legend. 
    }
    \label{fig:heatmap}
\end{figure}

After applying the prescription in \autoref{eq:Rmol04} to each galaxy, we compute the total \hi\ in each main halo (with contributions from the central and satellite galaxies) and present a bi-dimensional histogram of the \hi\ and dark matter masses of each halo in \autoref{fig:heatmap}. The solid, black line represents the running mean $M_{\hi}(M_h)$ relation as a function of halo mass ($M_{\text{h}}$). The dashed black line shows the equivalent result when using the prescription described by \autoref{eq:Blitz}. This figure can be understood as the halo occupation model predicted by the SAGE galaxies for \hi\ gas. We find that the prescription with $R_{\rm mol}=0.4$ predicts a larger total \hi\ mass around $M_h \simeq 10^{12}\,{\rm M}_{\sun}$ than the \citet{Obreschkow2009} prescription. However, both prescriptions tend to converge at the high- ($M_h \gtrsim 10^{13}\,{\rm M}_{\sun}$) and low-mass ($M_h \lesssim 10^{11}\,{\rm M}_{\sun}$) ends. 
In \appref{sec:fit} we show the split between central and satellite galaxy contributions, alongside with analytical fits to the $M_\hi (M_h)$ relations.

In the same figure, we compare our derived \hi--halo relation to other similar relations derived in the literature from \cite{padmanabhan} ($M_{\hi,1}$), \cite{bagla} ($M_{\hi,2}$), \cite{baugh} ($M_{\hi,3}$ and \cite{spinelli} ($M_{\hi,4}$)). Both \cite{baugh} and \citet{spinelli}  come from different simulation prescriptions with different implementations of (sub-grid) galactic physics. On the other hand, \cite{padmanabhan} matches the observational \hi\ luminosity function to a theoretical halo mass function, following a Halo Abundance Matching technique. A similar approach is followed by \citet{bagla}. 
The analytical expressions of these curves can be found in \appref{sec:analytical}.

Some general features are common to the shape of the different \hi--halo mass relations derived in the literature. On the low-$M_h$ end, $M_\hi$ follows a power-law with the halo mass. At the high-$M_h$ end, AGN are ---as their name implies--- very active, preventing hot gas from cooling, effectively suppressing the \hi\ abundance.  A detailed study on this relation is presented in \citealt{Chauhan2020} (see also references therein). That study also provides a fitting function of $M_\hi(M_h)$ as a function of redshift. We do not include that line in \autoref{fig:heatmap} to avoid overcrowding, but the shape is very similar to $M_{\hi,3}$.
In the  $M_{\hi,3}$ and  $M_{\hi,4}$ curves together with the two derived from SAGE galaxies, we see a second rise at higher masses due to the contribution of \hi\ from the satellites (see \appref{sec:fit}).

The variety of curves shown in \autoref{fig:heatmap} reflects the uncertainty within the literature about how the \hi\ fills dark matter haloes. These variations will lead to variations in the total abundance of \hi\ (discussed in \autoref{sec:omega} below), the large scale bias of \hi\ and the small-scale clustering of \hi. The latter may be relevant for \hi\ spectroscopic galaxy surveys, but should be negligible for the scales explored by \hi\ IM clustering. Both the abundance and bias of \hi\ will have an impact on the amplitude of the \hi\ IM clustering signal (see \autoref{sec:intensity_maps} below), but not on the shape of the 2PCF, which is the focus of Sections \ref{sec:HI_2PCF} \& \ref{sec:BAO}. 

The prescription described by \autoref{eq:Blitz} was only incorporated in this study at the final stages in order to better understand the \hi--halo relation and how it affects the \hi\ abundance. This technique may be more sophisticated than our original prescription based on \autoref{eq:Rmol04}. However, given that we do not expect any changes in the large scales studied in Sections \ref{sec:HI_2PCF} \& \ref{sec:BAO} and that \autoref{fig:heatmap} shows that the modelling uncertainty in $M_\hi (M_h)$ spans a large range of possibilities  beyond the differences between our two $R_{\rm mol}$ choices, we keep our original $R_{\rm mol}=0.4$ prescription as our default. We leave the study of the small-scale clustering of \hi\ emission-line galaxies when modelled with different prescriptions for future work. 

Likewise, a different calibration of \text{SAGE} or choosing $f_{\rm H}$ different to BBN could lead to a different shape or amplitude of $M_{\hi}(M_h)$, with negligible consequences for Sections \ref{sec:HI_2PCF} \& \ref{sec:BAO}. Hence, the results of this section needs to be interpreted as a net prediction of the \hi-halo relation for the \textsc{SAGE} \citep[as calibrated in][]{knebe18} applied to the UNITsims and a comparison with previous works.

\subsection{\hi\ abundance}
\label{sec:omega}

The abundance of \hi\ in the Universe across cosmic history, $\Omega_\hi(z)$, is poorly constrained and a source of debate. 
In fact, it is one of the most immediate topics that \hi\ intensity mapping is expected to shed light on, as it modulates the amplitude of the \hi\ IM clustering \citep{Masui:2012zc,Pourtsidou:2016dzn,GBTeBOSS}. 

In the previous subsection we described how we obtain the \hi\ mass for each galaxy in our simulation from the cold gas derived by SAGE. 
This immediately gives us a prediction of the total abundance of \hi\ by summing over all galaxies: 
\begin{equation}
\begin{split}
\label{eqn:omega1}
    \Omega_{\hi} = \frac{\rho_{\hi}}{\rho_{\rm crit}} = \frac{1}{\rho_{\rm crit}} \left(\frac{\sum_i M_{\hi,i}}{V_{\rm box}}\right) &= 3.8 \times 10^{-4}\,  \\
    &\{ = 2.7 \times 10^{-4} \} \, , \\
    \end{split}
\end{equation}
in which $\rho_{\rm crit}$ is the critical density of the Universe. For the top value, we have followed our default choice of $R_{\rm mol}=0.4$ (\autoref{eq:Rmol04}) and the bracketed value corresponds to using \autoref{eq:Blitz}. 

\begin{figure}
    \centering
    \includegraphics[width=0.48\textwidth]{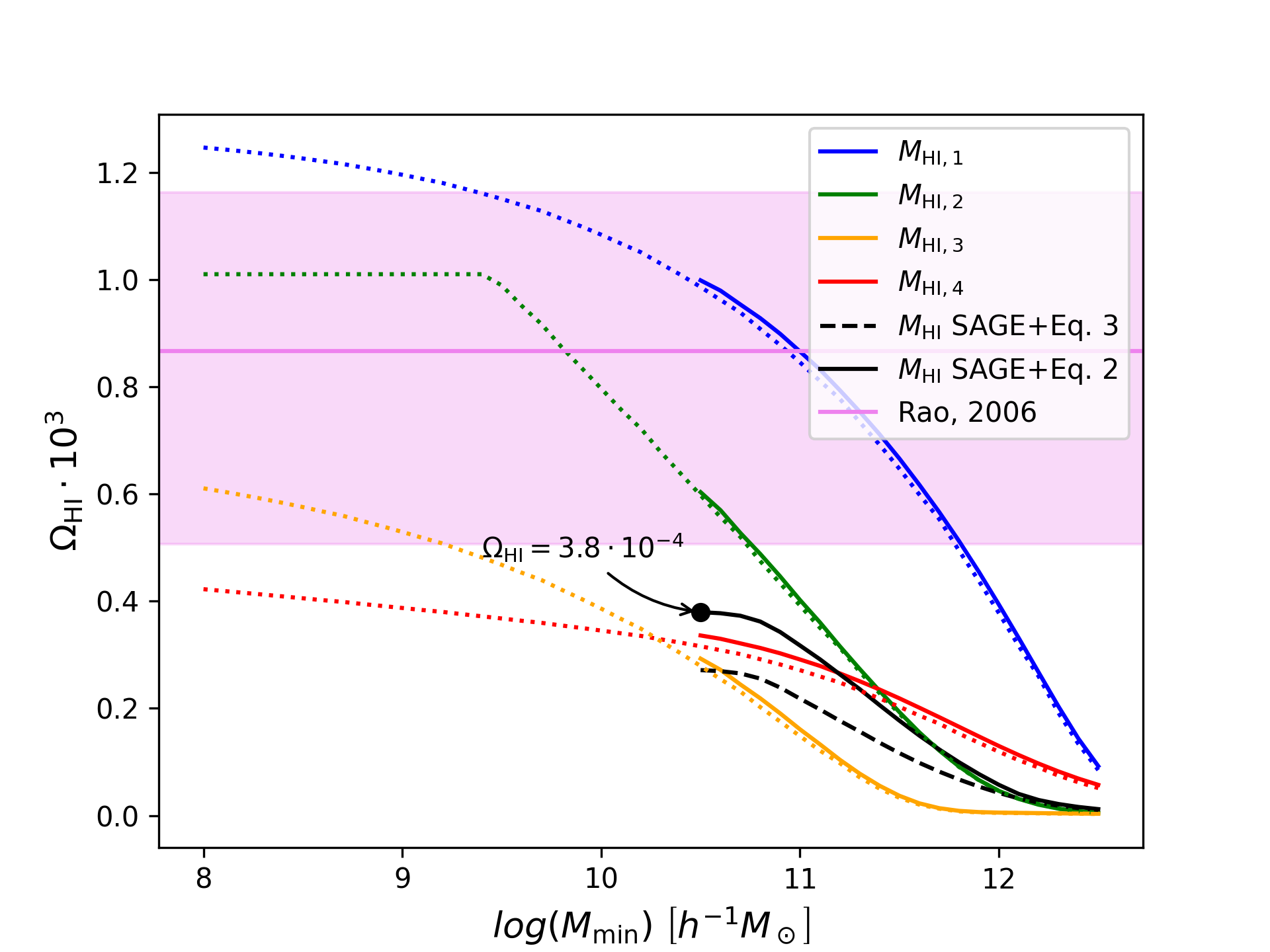}
    \caption{Inferred \hi\ density parameter from simulations as a function of the minimum halo mass considered. 
    The solid lines represent the different $M_\hi(M_h)$ prescriptions shown in \autoref{fig:heatmap} (same subscript notation) integrated with the halo mass function of the \unit\ (\autoref{eqn:omega2}), including the prescription derived from the SAGE galaxies and using $R_{\rm mol}=0.4$ (\autoref{eq:Rmol04}). The dashed black line represents the \hi\ obtained in \unit, when applying the \citet{blitz} prescription (\autoref{eq:Blitz}).
    The dotted lines are the same $M_\hi(M_h)$ prescriptions integrated with an analytical halo mass function (see text), for which we can reach lower masses. 
    The black point indicates the \hi\ density parameter obtained for \unit-SAGE galaxies for the resolution of the simulations and the default prescription ($R_{\rm mol}=0.4$). This is the value used later in this work in the  calculation of the brightness temperature in \secref{sec:HI_2PCF}. We also include as an horizontal band the $\Omega_\hi$ measurements from \citet{Rao2006} performed at redshift $1<z<1.5$.
    }
    \label{fig:omegas}
\end{figure}

One caveat for this prediction is that we are limited by the simulation's mass resolution. Although \autoref{fig:heatmap} shows that the \hi\ mass decays for lower halo masses, one needs to bear in mind that the halo mass function (HMF) grows rapidly for lower masses. In order to better understand  this effect, we can compute the 
\hi\ density parameter by integrating the HMF and \hi--halo mass relation from a given minimum mass $M_{\rm min}$:
\begin{equation}
\label{eqn:omega2}
    \Omega_{\hi}(M_{\rm min}) = \frac{1}{\rho_{\rm crit}} \int_{M_{{\rm min}}}^\infty \frac{\text{d}n(M_\text{h})}{\text{d}M_\text{h}} M_\hi(M_\text{h})\, \text{d}M_\text{h} \, .
\end{equation}
We now explore how the \hi\ density parameter is affected by the mass resolution in \autoref{fig:omegas}. The black lines show the $\Omega_\hi$ obtained from the UNITsims empirical HMF and \hi--halo relation (solid for \autoref{eq:Rmol04}, dashed for \autoref{eq:Blitz}) for different mass cuts. We recover the value reported in \autoref{eqn:omega1} for the case of $M_{\rm min} = 10^{10.5}\,h^{-1}\,{\rm M}_{\sun}$ (considered the halo mass resolution limit of the simulations, per \secref{sec:unitsim}). 

As discussed in \secref{sec:HIprescription}, there exists some theoretical uncertainty on the relation between the \hi\ mass and total mass of haloes. If we apply different $M_\hi(M_\text{h})$ prescriptions to the same simulation, using \autoref{eqn:omega2} we will naturally obtain a different $\Omega_\hi$. This is precisely what the other solid lines in \autoref{fig:omegas} represent.

Whereas for $M_{\hi,4}$ and the results derived from SAGE there are some hints of $\Omega_\hi$ converging already at $M_{\rm min} \simeq 10^{10.5}\,h^{-1}\,{\rm M}_{\sun}$, this is clearly not the case for $M_{\hi,1}$ and $M_{\hi,2}$. 
In order to avoid the inherent mass resolution limit of the UNITsims, we can use analytic HMF expressions  
computed with the public code \textsc{HMFCalc}\footnote{\url{https://hmf.icrar.org/}  \citep{murray}}. Following the reference associated to each prescription,  we use the \citet{ShethHMF} HMF for $M_{\hi,1}$ \citep{padmanabhan}, the \citet{Watson} HMF for $M_{\hi,2}$ (more adequate for FoF halos used in \citealt{bagla}) and the usual \citet{Tinker08} spherical overdensity main halo mass function for $M_{\hi,3}$ \& $M_{\hi,4}$ \citep{baugh,spinelli}.
Combining these HMF with the analytic expressions of the $M_\hi(M_\text{h})$ relation from those references, we show the \hi\ density parameter $\Omega_\hi(M_{\rm min})$ (dotted lines in \autoref{fig:omegas}) for $M_{\rm min}$ below the resolution of our simulations. 

As one would expect, for low halo masses, the four dotted lines converge to certain values, proving that there is no significant \hi\ missed from halo masses below $10^8\,h^{-1}\,{\rm M}_{\sun}$. The fast flattening of the $\Omega_{\hi,2}$ curve at $M_h=2.8 \times 10^{9}\,h^{-1}\, {\rm M}_{\sun}$ occurs because this prescription sets a minimum circular velocity that corresponds to that halo mass. 
The small differences we observe between the solid and dotted lines are a consequence of the differences between the cumulative mass function from the fitting formulas used and the one obtained directly from the \unit.

In this subsection, we have reported the value we obtain for $\Omega_\hi(z=1.321)$ in our simulation along with providing other estimates in \autoref{fig:omegas} that represent the theoretical side of the uncertainty on $\Omega_\hi$. \citet{Rao2006} set observational constrains on the abundance of \hi\ in the $1<z<1.5$ range to be $\Omega_{\hi} = 8.7^{+3.0}_{-3.6} \times 10^{-4}$, represented by the horizontal band in \autoref{fig:omegas}. Hence, our predictions given by \autoref{eqn:omega1} for the prescriptions in \autoref{eq:Rmol04} and \autoref{eq:Blitz} are $1.4\sigma$ and $1.7\sigma$ away from the observations, respectively, in no significant tension. Remarkably, all values reported in \autoref{fig:omegas} for $M_{\rm min}\leq 10^{10.5}\,h^{-1}\,{\rm M}_{\sun}$ are within the 2-$\sigma$ interval around the observed data.

Again, the $\Omega_\hi$ prediction may vary if the SAGE calibration or other parts of the pipeline were changed. Nevertheless, the impact of $\Omega_\hi$ on the \hi-IM clustering would simply be a change in the amplitude and would not affect the conclusions drawn later in Sections \ref{sec:HI_2PCF} \& \ref{sec:BAO}. The aim of this subsection was to estimate the theoretical uncertainty on $\Omega_\hi (z=1.321)$ and to study the minimum mass at which $\Omega_\hi(M_{\rm min})$ starts to converge. These estimates may be used in future \hi\ works, where the detailed knowledge of $\Omega_\hi$ may be crucial, for example when taking into account detector noise and estimating SNR for a given experiment. Additionally, we find that our SAGE-\unit\ prediction hints to a convergence, that it is, in general lines, in good agreement with other theoretical curves and that it is not in significant tension with existing data. Hence, these tests serve as a further step of the validation of the simulations used here.

\section{\hi\ Intensity Mapping anisotropic clustering}
\label{sec:HI_2PCF}

In this section, we outline how we create simulated Intensity Maps (IM) from the simulated \hi\ galaxies outlined in \secref{sec:simulation}, and analyse their clustering using isotropic and anisotropic two-point correlation functions (2PCF). Moreover, we simulate two observational effects that impact the \hi\ clustering: the telescope beam angular smoothing and foreground cleaning.

\subsection{Intensity maps}
\label{sec:intensity_maps}

\hi\ intensity mapping consists of integrating the entire signal of the redshifted 21-cm line in a given voxel of a chosen angular and frequency resolution. In our cubic simulation, we approximate those voxels by cubic pixels\footnote{Here, both pixel and voxel refer to 3D cells. The former with cubic Cartesian limits, the latter with spherical caps and angular cuts as their limits.} of side $l=5\,h^{-1}\,{\rm Mpc}$. This is the typical size of pixels along the angular direction in this kind of analysis \citep[e.g.][]{Wang:2020lkn}, although in reality the radial resolution is much greater; that is, we would expect a smaller $l_z$. For this paper, we assume that the voxels are rebinned in the radial direction in order to obtain a similar resolution perpendicular and parallel to the line of sight (for example, \citealt{GBTeBOSS} used $l_z \simeq 5\,h^{-1}\,{\rm Mpc}$).
In practice, this means 
summing the \hi\ mass of all the galaxies ($j$)  within each cell $i$:
\begin{equation}
    M_{\hi ,{i} }= \sum_{j \in \,\text{cell} \hspace{1mm}i} M_{\hi ,j} \hspace{2mm} .
\end{equation}
This is equivalent to a Nearest-Grid-Point approach with weights proportional to the \hi\ mass of each galaxy. In order to study the \hi\ clustering, one would compute the \hi\ mass overdensity $\delta_{\hi}$ in each pixel:
\begin{equation}
    \delta_{\hi ,i} = \frac{M_{\hi ,i}}{\bar{M}_\hi}-1 \, ,
\end{equation}
with $\bar{M}_\hi$ being the average \hi\ mass over all pixels. 
However, the observable for intensity mapping experiments is temperature variation $\Delta T$, which is directly proportional to mass variation $\Delta M_i = M_{\hi ,i} - \bar{M}_\hi$. For simplicity, the total mass of \hi\ is absorbed into the mean brightness temperature $\bar{T}_\text{b}$ via the $\Omega_\hi$ parameter \citep{Battye:2012tg}:  
\begin{equation}
    \bar{T}_\text{b}(z) = 190 \, \frac{H_0 \left(1+z\right)^2}{H(z)}\Omega_\hi(z) \,h = 1.820 \times 10^{-1} \, \text{mK}
\end{equation}
so the final intensity maps are expressed as a function of the \hi\ overdensity: 
\begin{equation}
    \Delta T_i = \bar{T}_\text{b} \delta_{\hi,i} .
\end{equation}
Here, we use the $\Omega_\hi$ derived from the \unit-SAGE catalogues (\autoref{eqn:omega1}).

\begin{figure*}
    \centering
    \includegraphics[trim=220 60 200 60, clip, width=0.8\linewidth]{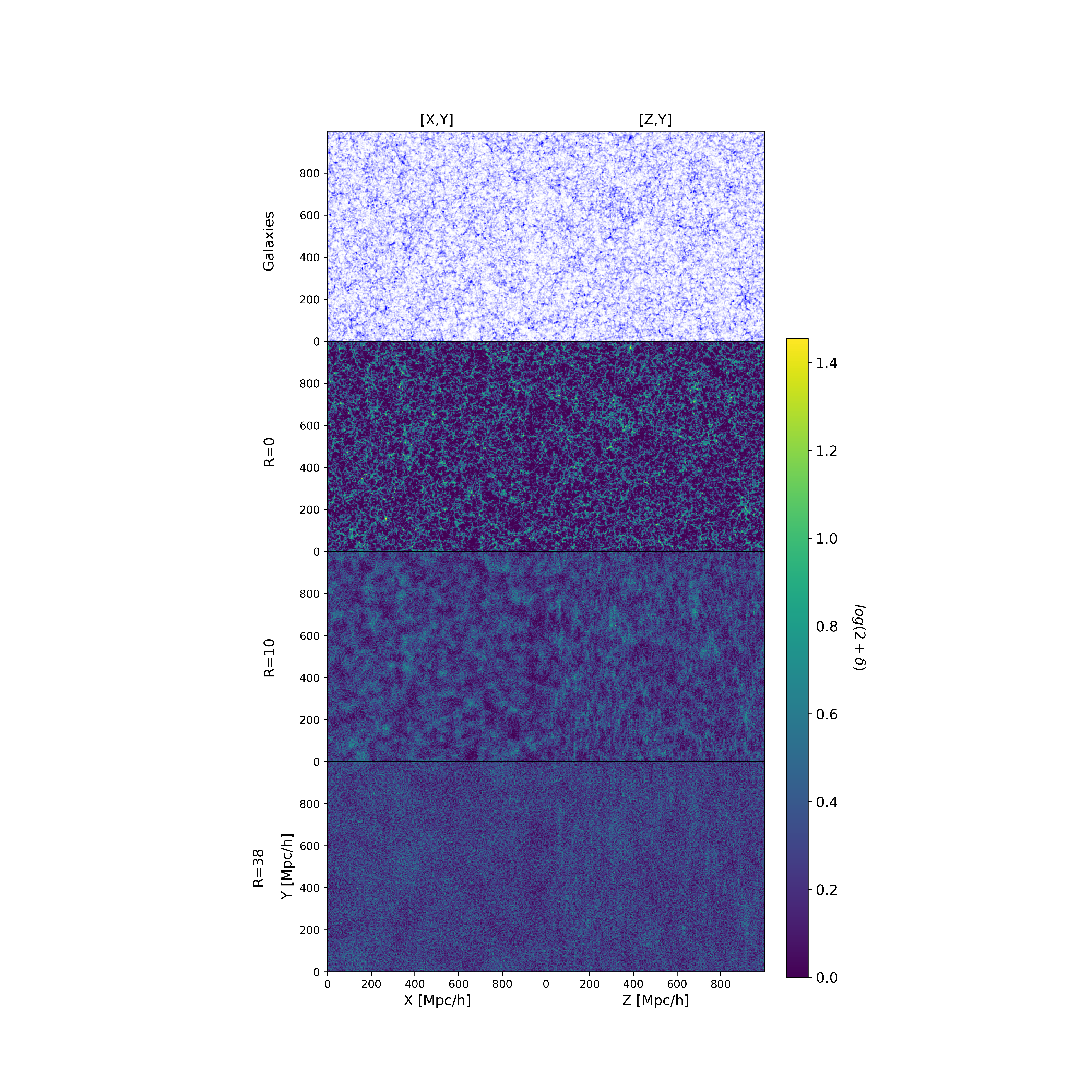}
    \caption{We show two 5 Mpc$/h$-deep slices of one of the simulation boxes (UNITSIM1): one showing the $[X,Y]$ plane (left) and one showing the $[Z,Y]$ plane (right). On the top panels we show the position of the galaxies with a size proportional to their \hi-mass. On the second row, we show the pixelised \hi\ intensity maps. On the third and fourth rows, we show the same intensity maps after applying an angular smoothing on the $[X,Y]$ plane with length $R=10 {\rm Mpc}/h$ and $R=38{\rm Mpc}/h$, respectively.}
    \label{fig:maps}
\end{figure*}

In order to visualise the intensity mapping technique, we show on the top row of \autoref{fig:maps} the galaxy positions in one slice of the simulation box and, on the second row, the equivalent pixelised \hi\ intensity map.

\subsection{Simulating the angular beam}
\label{sec:beam}

The intensity mapping technique inherently implies an angular smoothing due to the telescope beam. For the case of an SKA1-MID 21-cm intensity mapping program in single-dish mode, we obtain a smoothing angle of
\begin{equation}
    \theta_\text{FWHM} = \lambda/D_\text{dish}  \simeq  0.8 (1+z) \, \text{deg}
\end{equation}
\citep[see e.g.][]{villaescusa}, where $\lambda$ is the wavelength of the observation and $D_\text{dish}$ the diameter of the antenna's dish. This translates to a comoving scale given by 
\begin{equation}
        R_{\rm beam} = \frac{r(z) \sin\left(\theta_\text{FWHM}\right)}{2\sqrt{2 \ln(2)}} =  38.45\, h^{-1} \text{Mpc}
\end{equation}
at $z=1.321$. In later sections, we also consider an intermediate case of $R_{\rm beam}=10\,h^{-1}\, {\rm Mpc}$, which may be interpreted as a lower-redshift study or an observation with a larger dish.

We assume that the telescope beam can be approximated to first order by a Gaussian convolution. In order to apply this to the simulations, we assign one of the axes of each box as the radial direction ($\vec X_{\parallel}$)%
\footnote{Even though $\vec X_\parallel $ is one-dimensional, we add the vector symbol to distinguish displacement vectors ($\vec r_\parallel$) from their modulus $r_\parallel=\lvert \vec{r}_\parallel \lvert $.}
and the other two as the `angular' axes ($\vec X_{\perp}$). We then move the galaxies along the angular axes with a 2D-Gaussian random ($\mathcal{N}$) number with a variance $\sigma^2 = R_{\rm beam}^2$:
\begin{equation}
\begin{split}
    \vec X_{\parallel}^{\rm smooth}  &= \vec X_{\parallel} \\
   \vec X_{\perp}^{\rm smooth}  &\curvearrowleft \mathcal{N}\Big(\mu = \vec X_{\perp}, \sigma= \{ R_{\rm beam}, R_{\rm beam}\}\Big) 
   \end{split}
\end{equation}
Where $\{\vec X_{\perp}^{\rm smooth},X_{\parallel}^{\rm smooth}\}$ represent the new coordinates of a given galaxy. 
When including the angular beam, this step must be performed prior to the pixelisation described in \secref{sec:intensity_maps}. 

Alternatively, we also tried a direct 2D-Gaussian convolution to the intensity maps in order to incorporate the angular beam, obtaining identical results. However, we found that method less computationally efficient  than the particle-based method explained above. 

We can visualise on the last 3 rows of \autoref{fig:maps} how increasing the telescope beam affects the intensity maps. We see that on the $[X,Y]$ plane, which is considered the angular one, the structures are significantly smoothed out. On the $[Y,Z]$ plane, we see again the structures being smoothed along the $Y$-axis, but remaining along the $Z$ (radial) axis.

\subsection{Simulating foreground removal}
\label{sec:foregrounds}

One of the challenges of performing precision cosmology with IM is the presence of foregrounds that emit signals at the same observer-frame frequency as \hi. One example is the Galactic synchrotron radiation produced by the radial acceleration of charged relativistic particles which can have a flux several orders of magnitude greater than the \hi\ \citep{Alonso:2014sna,cunnington19}. 

The most robust techniques to remove the foregrounds are based on the knowledge that foregrounds emit a spectrum that is smooth in frequency, whereas the cosmological clustering of \hi\ provides an fluctuating component to the total observed spectrum. However, the largest-wavelength Fourier modes of the cosmological signal are interpreted as a smooth signal and are removed by these methods. This results in an effective scale $2\pi/k_{\rm FG}$ above which the observed cosmological signal is exponentially suppressed along the radial direction (e.g. $Z$-axis). Following \citet{soares}, this scale depends  both on the foreground removal method itself (characterised by $N_\parallel$) and the radial length of the survey considered ($L_z$): 
\begin{equation}
    k_{\rm FG} = N_{\parallel} \frac{2\pi}{L_z} \, .
\end{equation}
We consider $N_\parallel= 2$, approximately emulating a FastICA method with $N_\text{IC}=4$ independent components removed. We consider two cases with $L_z=3450\,h^{-1}\,{\rm Mpc}$ and $L_z=300\,h^{-1}\,{\rm Mpc}$, corresponding to the full SKA range $0.35<z<3$ and a small redshift bin $1.1<z<1.3$,  giving $k_{\rm FG}=3.64 \times 10^{-3}\,h\,{\rm Mpc}^{-1}$ and $k_{\rm FG}=4.19 \times 10^{-2}\,h\,{\rm Mpc}^{-1}$, respectively.

In order to apply this effect to our grid of temperature fluctuations, we first apply a 1D fast Fourier transform to each radial line of pixels:
\begin{equation}
    \widetilde{\Delta  T}_{lm}(k_\parallel) = \sum_{r_\parallel} \Delta T_{lm}(r_\parallel)\, \exp\left( \frac{-2 \pi \,i \,r_\parallel k_\parallel}{N}\right) \hspace{3mm} \forall l,m \in \left\lbrace 1,...,N\right\rbrace \,,
\end{equation}
in which indices $l,m$ run over the angular plane (e.g. $\{X,Y\}$) and $N=200$ is the number of grid points along each direction. We omit these indices in later equations for simplicity.

Once we obtain $\widetilde{\Delta T}(k_\parallel)$, we can attenuate the largest scales using a transfer function fitted in \cite{soares} to full foreground + foreground-removal simulations:
\begin{equation}
     f (k_\parallel) = 1 - \exp\left[ -\left(\frac{k_\parallel}{k_{\rm FG}}\right)^2\right] \, .
\end{equation}
After multiplying by the transfer function, we transform back to configuration space to obtain the attenuated intensity maps: 
\begin{equation}
    \Delta T^\text{FG}_{lm} (r_\parallel) = \sum_k \widetilde{\Delta  T}_{lm}(k_\parallel) \, f(k_\parallel) \, \exp\left( \frac{2 \pi i r_\parallel k_\parallel}{N} \right) \, .
\end{equation}

\subsection{HI Clustering}
\label{sec:2PCF}
\paragraph*{Isotropic 2PCF $\xi(r)$}
\ \\

\noindent In the previous subsections we described how we build intensity maps from our galaxy catalogues and add the observational effect of the angular beam and the foreground subtraction process.  We now analyse the 2-point correlation function of these intensity maps, which is defined as
\begin{equation}
\label{eq:2PCF}
    \xi (\lvert \vec r \lvert) = \left\langle \Delta T (\vec X)\, \Delta T (\vec X + \vec r) \right\rangle\,, 
\end{equation}
where $\langle \cdot \rangle$ represents the average of all positions $\vec X$ and all orientations of $\vec r$. In practice, the average is performed over all pairs of pixels separated by a distance $r = \lvert \vec r \lvert$, within a bin width ($r_i-\Delta r /2<r< r_i + \Delta r/2$) that we set to $\Delta r= l_{\rm cell}=5\,h^{-1}\,{\rm Mpc}$.

\paragraph*{Anisotropic 2PCF $\xi(r_\perp, r_\parallel)$}
\ \\

\noindent Even though this study focuses on real-space clustering (hence we do not expect any cosmological anisotropy), the observational effects described in \secref{sec:beam} \& \ref{sec:foregrounds}
will affect the radial and angular distances differently. For that reason, we will put the focus of this work on the anisotropic 2-point correlation function, defined in an analogous way: 
\begin{equation}
\label{eq:2PCF_2D}
    \xi (| \vec r_\perp |, | \vec r_\parallel |) = \left\langle \Delta T ( \vec X_\perp , \vec X_\parallel)\cdot  \Delta T ( \vec X_\perp + \vec r_\perp , \vec X_\parallel + \vec r_\parallel )  \right\rangle \, .
\end{equation}
The average  $\langle \cdot \rangle$ is still performed over all pairs of pixels, but we now bin the results in two distances $ r_\perp = | \vec r_\perp |$, $ r_\parallel = | \vec r_\parallel |$ with the same bin width $\Delta r$.
We ran the anisotropic 2PCF over the four UNITsims boxes, and consider three different orientations in which each axis ($X,Y,Z$) is chosen as the radial one for each case. This gives us 12 different 2PCFs, over which we average for the remaining figures in this paper. In some occasions, we also show error bars corresponding to the standard deviations of the 12 2PCFs, but we note that these error bars are not representative of any experiment. This is due to the different orientations not being completely independent, but also to the nature of the UNITsims that are constructed with reduced variance and paired initial conditions (see discussion in \secref{sec:unitsim}). However, the relative size of the error bars of different estimators can give us an idea of how much signal is lost by applying certain geometrical cuts or projections. 

\begin{figure*}
    \centering
    \includegraphics[trim = 32 0 60 0, width=1.0\textwidth]{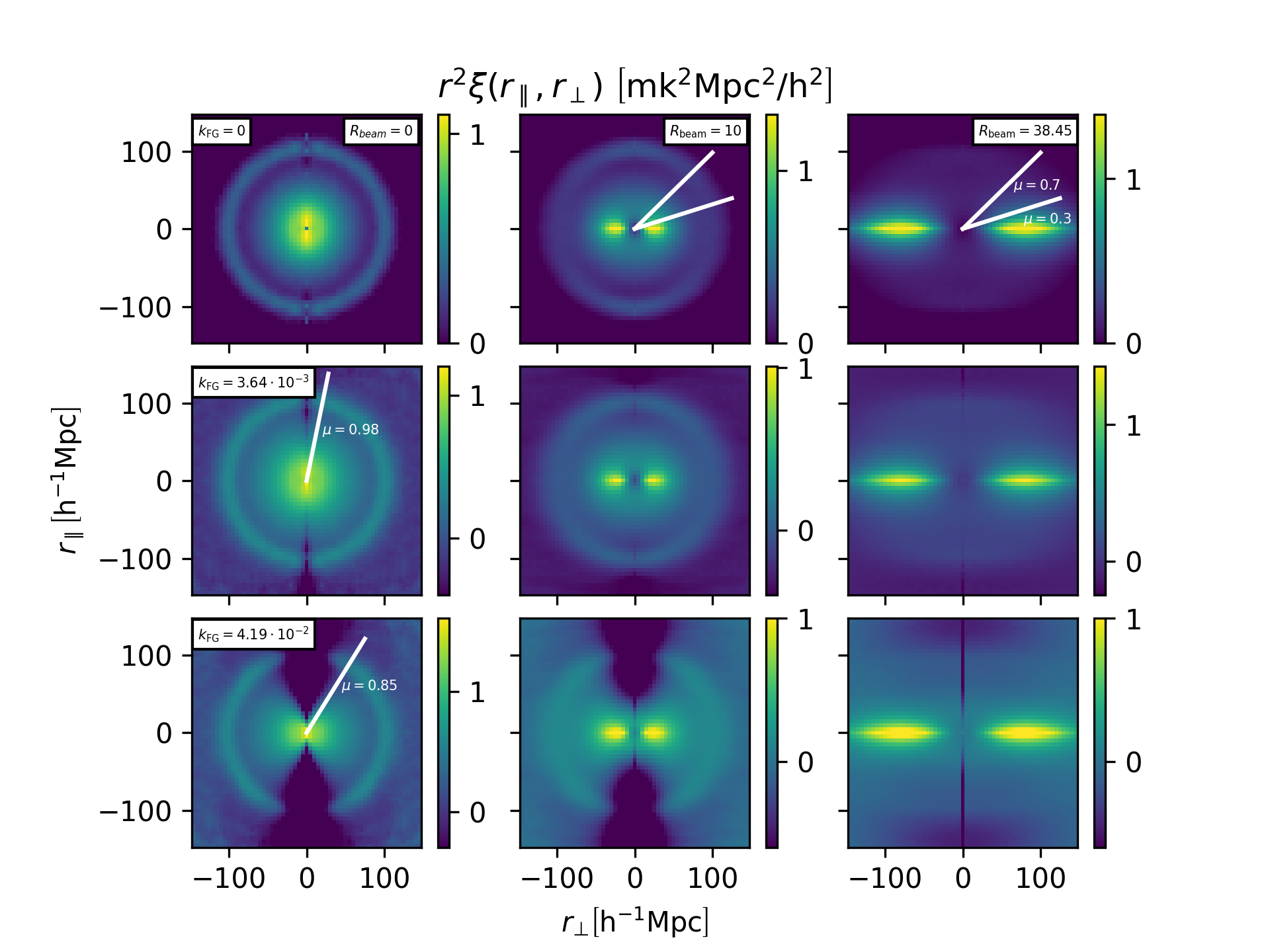}
    \caption{ 
    Anisotropic two-point correlation functions $\xi(r_\parallel,r_\perp)$ considering all nine possible combinations of the telescope beam and the foreground-removal effects considered in this work:  $R_{\rm beam}= \{ 0, 10, 38\}\,h^{-1}\,\text{Mpc}$\ $\otimes$ $k_{\rm FG} = \{ 0, 3.64, 41.9 \} \cdot 10^{-3}\,h\,{\rm Mpc}^{-1}$. The 2PCF is multiplied by $r^2$ to highlight the large scales. The four quadrants of each subplot are identical mirrored replicas, but are found very helpful for visualising the BAO and other features, and are commonly found in the literature when representing the  $\xi(r_\parallel,r_\perp)$ functions. The white lines mark limits in $\mu=r_\parallel/r$ considered in \secref{sec:BAO}.}
    \label{fig:2PCF2D}
\end{figure*}

In \autoref{fig:2PCF2D} we can see the nine anisotropic two-point functions obtained in this work, for the combinations with $R_{\rm beam} = \{0, 10, 38\}\,h^{-1}\,{\rm Mpc}$ and $k_{\rm FG} = \{0, 3.64, 41.9 \} \times 10^{-3}\, h\,{\rm Mpc}^{-1}$.
On the top left panel, we show the 2PCF for the purely cosmological case (without telescope beam or foreground-removal effects), where BAO can be seen as a ring at $r = (r_\perp^2+r_\parallel^2)^{1/2} \simeq r_{\rm BAO} \simeq 103\,h^{-1}\,\text{Mpc}$.
We find the signal to be almost entirely isotropic, with only some small artifacts along the $r_\perp=0$ axis, which likely come from a grid effect. We note that, along that axis, there are fewer pairs of pixels to average over, as angular distances are computed in 2D, whereas the radial one is one-dimensional.

The other two plots in the top row show the 2PCF for the cases where we have applied a mild ($R_{\rm beam} = 10\,h^{-1}\, {\rm Mpc}$) and a large ($R_{\rm beam} = 38\,h^{-1}\, {\rm Mpc}$) telescope beam. We observe how the signal is progressively smoothed along the $r_\perp$-axis for larger $R_{\rm beam}$, affecting all $\xi (r_\perp, r_\parallel)$ values, making the BAO feature dimmer and with two strong lobes appearing along the $r_\parallel=0$ axis. We additionally mark two orientations at $\mu=0.7$ and $\mu=0.3$ (see \autoref{eq:polar}) that will be used later as a $\mu_{\rm min}$-cut in order to isolate the BAO.

In a similar way, the two bottom panels of the left column of \autoref{fig:2PCF2D} represent the effect of foreground cleaning, parametrized by $k_{\rm FG}$. 
The middle panels have $k_{\rm FG} = 3.64\times 10^{-3}\,h\,\text{Mpc}^{-1}$, which is likely to be closer to the expected value for an SKA MID-1 observation, whereas the lower-left plot considers a more pessimistic value of $k_{\rm FG} = 4.19\times 10^{-2}\,h\, {\rm Mpc}^{-1}$. As we can see, the effect is much more localized and only affects a wedge of scales around the purely radial axis ($r_\perp \sim 0$), strongly suppressing the 2PCF, but leaving most of the plane intact. 
As expected, the effect becomes stronger for larger $k_{\rm FG}$ and we consider an orientation $\mu$, represented by the white lines, that marks the region most affected, as we will see in \autoref{sec:wedge}.

Finally, the four remaining panels in \autoref{fig:2PCF2D} consider the combined effect of both contributions, where we can see how both effects couple together. When $R_{\rm beam}=10\,h^{-1}\,{\rm Mpc}$, all plots are similar to $R_{\rm beam} = 0$, with a slightly more blurred BAO feature. For the largest beam, $R_{\rm beam} = 38\,h^{-1}\,{\rm Mpc}$, all the scales are strongly blurred, and the BAO feature becomes very subtle.

\autoref{fig:2PCF2D} is a good starting point for analysing BAO recovery, as it demonstrates observational effects separately in each dimension.
Equipped with the intuition that the BAO peak is still there, even for the strongest observational effects, in \secref{sec:BAO} we will analyse two families of methods (radial and $\mu-$wedges) to isolate the feature. 

\paragraph*{2PCF Multipoles $\xi_l$}
\ \\

\noindent First of all, we remind the reader that all the anisotropy studied here comes from the observational effects of the foreground removal and the telescope beam; that is, we have not applied redshift-space distortions or Alcock--Paczynski effects \citep{AP1979}. It is important to quantify these observational effects that are intrinsic to \hi\ intensity mapping so that we are aware of them when we compare anisotropic clustering analysis of \hi\ IM to spectroscopic galaxy survey analysis \citep{Blake:2019ddd,cunnington20}.

As a standard procedure, we expand the anisotropic correlation function $\xi(r_\parallel,r_\perp)$ in an orthogonal basis of Legendre polynomials:

\begin{equation}
    \xi_\ell (r) = \left(2\ell+1\right) \sum_\mu \xi\big( r_\perp (r,\mu),r_\parallel (r,\mu)  \big)\, \mathcal{L}_\ell(\mu)\, \Delta \mu \, ,
    \label{eq:multipoles}
\end{equation}
in which we use $\Delta \mu = 1/32$, where we have checked that convergence is reached. The conversion from Cartesian ($r_\parallel,r_\perp$) to polar coordinates  ($r, \mu$) is done with 
\begin{equation}
    \label{eq:polar}
     r = \sqrt{r_\parallel^2+r_\perp^2}\,,\quad \mu = \frac{r_\parallel}{r}\,. 
\end{equation}
For completeness, we note that $\mu$ is the cosine of the angle $\theta=\arctan(r_\parallel/r_\perp)$ with respect to the line of sight. Whereas we will not use this variable here, it is sometimes referenced in the literature. 
For the case of \hi~IM with foreground removal and a telescope beam, following the theoretical model in \citet{cunnington20,soares}, all even multipoles would be non-vanishing, but with decreasing relevance for higher $\ell$.
Here, we will consider the monopole ($\ell=0$), the quadrupole ($\ell=2$) and the hexadecapole ($\ell=4$), which are the ones broadly studied in the literature, since they are the only non-zero multipoles for galaxy clustering under the Kaiser approximation \citep[][]{Kaiser}. The expressions for the three first even Legendre polynomials are: 

\begin{equation}
     \mathcal{L}_0 (\mu) = 1 \,; \,\, \,\,
     \mathcal{L}_2 (\mu) = \frac{3 \mu^2 -1}{2} \,; \,\, \,\,
     \mathcal{L}_4 (\mu) = \frac{35\mu^4 -30\mu^2 +3}{8} \,. 
\end{equation}

\begin{figure*}
    \centering
    \includegraphics[width=1.04\textwidth]{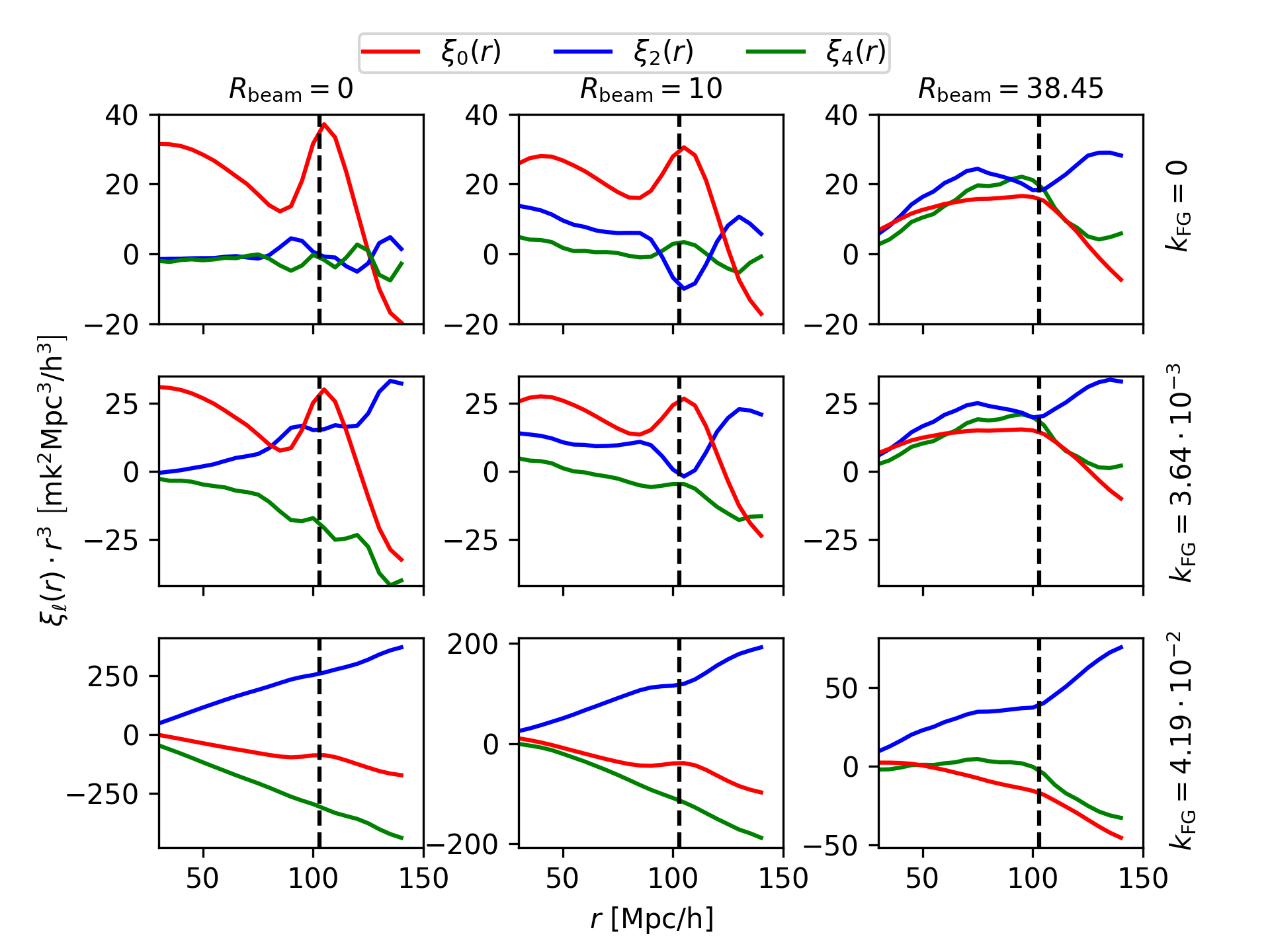}
    \caption{Representation of the first three even multipoles: monopole $\xi_0 (r)$, quadrupole $\xi_2 (r)$ and hexadecapole $\xi_4 (r)$. The 2PCF is multiplied by $r^3$ to highlight the large scales. We show all nine possible combinations of the telescope beam and foreground-removal effects considered in this work: $R_{\rm beam}= \{ 0, 10, 38\}h^{-1}\,{\rm Mpc}$\ $\otimes$ $k_{\rm FG} = \{ 0, 3.64, 41.9 \} \cdot 10^{-3}\, h\, {\rm Mpc}^{-1} $. The vertical dashed line represents the comoving size of the BAO $r_{\rm BAO}=103 h^{-1}\, {\rm Mpc}$ expected for this cosmology.}
    \label{fig:multipoles}
\end{figure*}

All multipoles we consider are represented in \autoref{fig:multipoles}, following the same structure in columns and rows as \autoref{fig:2PCF2D}.
For the case where neither the telescope beam nor the foreground-removal effects are considered, the signal is isotropic and, hence, the quadrupole and hexadecapole are consistent with $\xi_2 = \xi_4 =0$. As we increase either of the effects, the signal starts propagating to the other multipoles. Remarkably, a BAO signal can also be appreciated in the quadrupole for most cases, and sometimes the hexadecapole too. This is caused by cosmological information being propagated from the monopole to higher multipoles due the mixing of modes performed by the angular smoothing.

As a remark, the monopole represented in \autoref{fig:2PCF2D} and computed with \autoref{eq:multipoles} is equivalent to the isotropic 2PCF computed with \autoref{eq:2PCF} besides pixelisation effects, which are negligible at the scales of interest. 
The effect of the beam on the monopole has been studied also in \citet{villaescusa} and
\citet{Knennedy&Bull}, finding similar results. 
The recent work of \citet{Knennedy&Bull} also studies the effect of both $R_{\rm beam}$ and $k_{\rm FG}$ in the monopole and quadrupole. The latter results are similar to what we present in \autoref{fig:2PCF2D} for the monopole, whereas there are obvious differences for the quadrupole, but these are expected since that work is done in redshift space. Nevertheless, one common finding is that there is BAO signal in $\ell>0$ multipoles. A more direct comparison should be done in future works.

\section{Searching for the BAO signal}
\label{sec:BAO}

In this section, we study several methods to isolate the BAO signal that was seen in then anisotropic 2PCF $\xi(r_\perp,r_\parallel)$ in \secref{sec:HI_2PCF} . 

\subsection{$\mu$-wedge 2PCF}
\label{sec:wedge}

As we already hinted in the previous section, given the circular shape that is seen in \autoref{fig:2PCF2D}, even for the strong cases of observational effects, one natural way to isolate BAO would be to compute a $\mu$-wedged 2PCF. This means we average over bins with the same $r=\sqrt{r_\perp^2+r_\parallel^2}$ and discard measurements with values of $\mu$ (see \autoref{eq:polar}) heavily affected by the aforementioned effects. 
For that, we define the $\mu$-wedge 2PCF as
\begin{equation}
\label{eq:muwedge}
    \xi_{\mu_{\rm min} <\mu<\mu_{\rm max}} (r_i) = \frac{1}{N_i} \sum_{ \substack{r_i - \frac{\Delta r}{2} <r< r_i + \frac{\Delta r}{2} \\ 
    \mu_{\rm min} <\mu<\mu_{\rm max}} } \xi (r_\perp, r_\parallel) \hspace{3mm} ,
\end{equation}
where $N_i$ is the number of $\xi(r_\perp,r_\parallel)$ pixels that meet the $\mu$ cuts and belong to the $r$ bin with center at $r_i$ and width $\Delta r$. Note that in the case with $\mu_{\rm min}=0$ and $\mu_{\rm max}=1$ we would recover the monopole.

\paragraph*{$\bmath{\mu_{\rm min}}$ for the telescope beam}
\ \\

\begin{figure}
    \centering
    \includegraphics[width=0.52\textwidth]{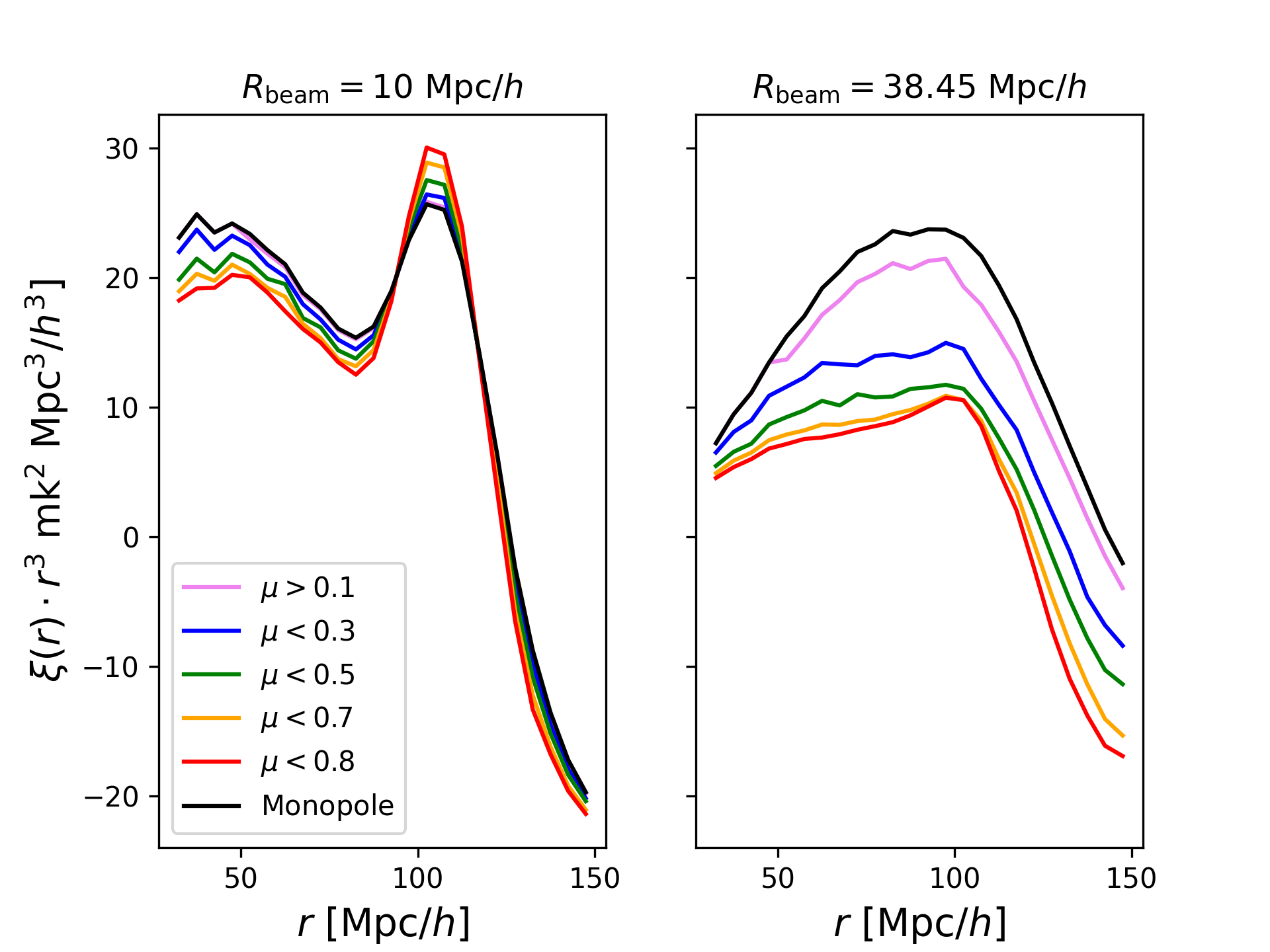}
    \caption{$\mu$-wedge 2PCF for a fixed $\mu_{\rm max}=1$ and a varying $\mu_{\rm min}$ as indicated in the legend. The left-hand panel takes $R_{\rm beam} = 10 \,h^{-1}\,{\rm Mpc}$, while the right-hand panel takes $R_{\rm beam} = 38\,h^{-1}\,{\rm Mpc}$. In particular, we select $\mu > 0.3$ and $\mu > 0.7$ for a more detailed analysis. See the angular cuts marked as white solid lines in  \autoref{fig:2PCF2D}.} 
    \label{fig:ringmethod}
\end{figure}

\noindent We first consider the effect of the telescope beam in isolation (i.e. $k_{\rm FG}=0$) for  $R_{\rm beam} = 10\,h^{-1}\,{\rm Mpc}$ and $R_{\rm beam} = 38\,h^{-1}\,{\rm Mpc}$. Given that the telescope has a greater effect when $\mu$ is low (i.e. small $r_\parallel$, thus large $\theta$), we do not consider an upper cut in $\mu$ (i.e. $\mu_{\rm max}=1$), but we let $\mu_{\rm min}$ vary and demonstrate the effect this has on results in \autoref{fig:ringmethod}. 

For the $R_{\rm beam} = 10\,h^{-1}\,{\rm Mpc}$ case, we find a clear BAO signal that gets sharper as we increase the angular cut. This means that the more stringent the constraint we place on $\mu$, the more we reduce the effects from the beam. For the larger telescope beam ($R_{\rm beam} = 38\,h^{-1}\,{\rm Mpc}$), similar results are found, but the BAO peak is subtler, as the shape of the 2PCF has changed significantly. 
For this case, although no convergence is reached, the shape and amplitude start to show some stabilisation around $\mu_{\rm min}=0.7$.
One could be tempted to put a very high $\mu_{\rm min}$ to obtain a more prominent BAO, but this means removing a large amount of the signal and hence increasing the size of the error bars, as we will see in the following subsections.
In the remainder of this paper we will consider $\mu_{\rm min} = 0.7$ and $\mu_{\rm min}=0.3$ to probe both the case with a more prominent peak and the case with smaller uncertainty on the 2PCF. These $\mu_{\rm min}$ cuts are shown by the white lines in \autoref{fig:2PCF2D}.

\paragraph*{$\bmath{ \mu_{\rm max}}$ for foregrounds}
\ \\

\noindent The foreground-removal method damps the signal when close to the line-of-sight, as can be seen in \autoref{fig:2PCF2D}. In order to perform an analysis analogous to the previous subsection, we take $\mu < \mu_{\rm max}$ cuts, avoiding the wedge near the radial axis ($r_\perp \simeq 0$, $\mu \simeq 1$). 

\begin{figure}
    \centering
    \includegraphics[width=0.52\textwidth]{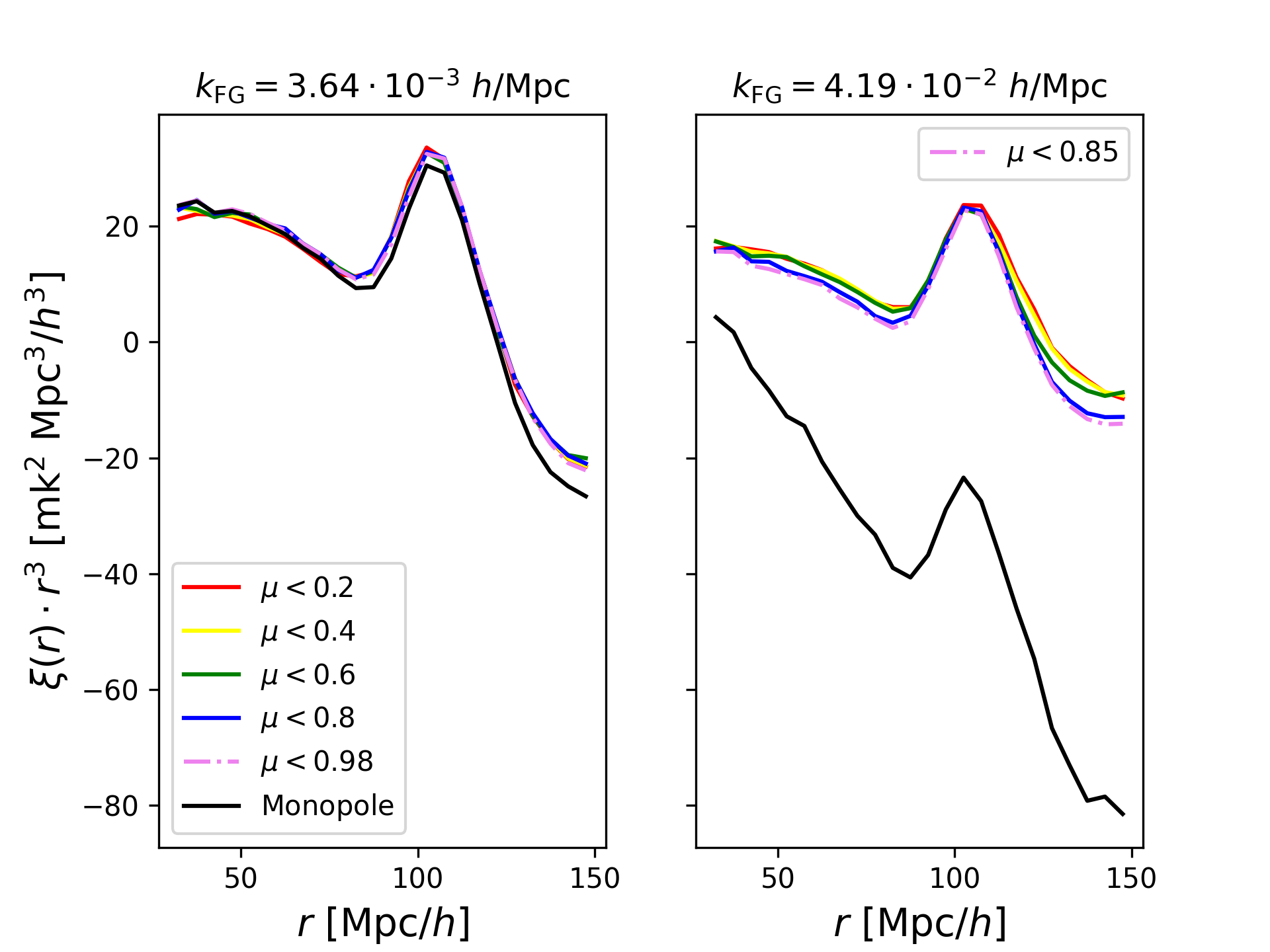}
    \caption{$\mu$-wedge 2PCF for a fixed $\mu_{\rm min}=0$ and varying $\mu_{\rm max}$ as indicated in the legend. On the left we consider $k_{\rm FG} = 3.64 \times 10^{-3}$ $\,h\, {\rm Mpc}^{-1}$, while on the right $k_{\rm FG} =4.19 \times 10^{-2}\,h\, {\rm Mpc}^{-1}$. The violet dash-dotted curves show the minimal $\mu_{\rm max}$ that is able to remove the foreground-cleaning effect. This optimal $\mu_{\rm max}$ is tuned for each $k_{\rm FG}$ and they are marked as white solid lines in  \autoref{fig:2PCF2D}.}
    \label{fig:ringmethodforegrounds}
\end{figure}

In \autoref{fig:ringmethodforegrounds} we represent the recovered $\mu$-wedge correlation functions considering  $k_{\rm FG} = 3.64 \times 10^{-3}$ and $k_{\rm FG} = 4.19 \times 10^{-2}$ $h\, {\rm Mpc}^{-1}$ for $R_{\rm beam}=0$, for different $\mu_{\rm max}$ and a fixed $\mu_{\rm min}=0$. As we can see, the BAO signal is always recovered regardless of $\mu_{\rm max}$. The effect of the choice of $\mu_{\rm max}$ is generally small, with the exception that the monopole is found offset (especially for $k_{\rm FG}=4.19 \times 10^{-2}\,h\,{\rm Mpc}^{-1}$) with respect to the rest of curves. This is because the negative contribution from the $\mu \simeq 1$ region drags all the monopoles down. 
We searched for the minimal $\mu_{\rm max}$ that is able to remove this effect, finding $\mu_{\rm max}=0.98$ and $0.85$ for  $k_{\rm FG} = 3.64 \times 10^{-3}$ and $k_{\rm FG} = 4.19 \times 10^{-2}\,h\, {\rm Mpc}^{-1}$, respectively. These $\mu$-cuts are represented as white lines in \autoref{fig:2PCF2D}, where they are shown to clearly avoid the most obviously foreground-affected regions. The 2PCF wedges with the optimal $\mu_{\rm max}$ are represented by the violet dot-dashed line in \autoref{fig:ringmethodforegrounds}, showing that these cuts are sufficient to de-bias most of the foreground effect.
 Nevertheless, in terms of BAO recovery it is not clear at this stage the necessity to perform any upper cut on $\mu$.

For the strong foreground case ($k_{\rm FG}=4.19\times 10^{-2}\,h\,{\rm Mpc}^{-1}$) we also notice that the BAO becomes slightly sharper as we put tighter cuts on $\mu$. However, this latter effect is very mild effect and unlikely to be beneficial if one considers the loss of signal.

\paragraph*{$\bmath{ \mu_{\rm min}<\mu<\mu_{\rm max}}$ for combined effects}
\ \\

\begin{figure*}
    \centering
    \includegraphics[width=1.06\textwidth]{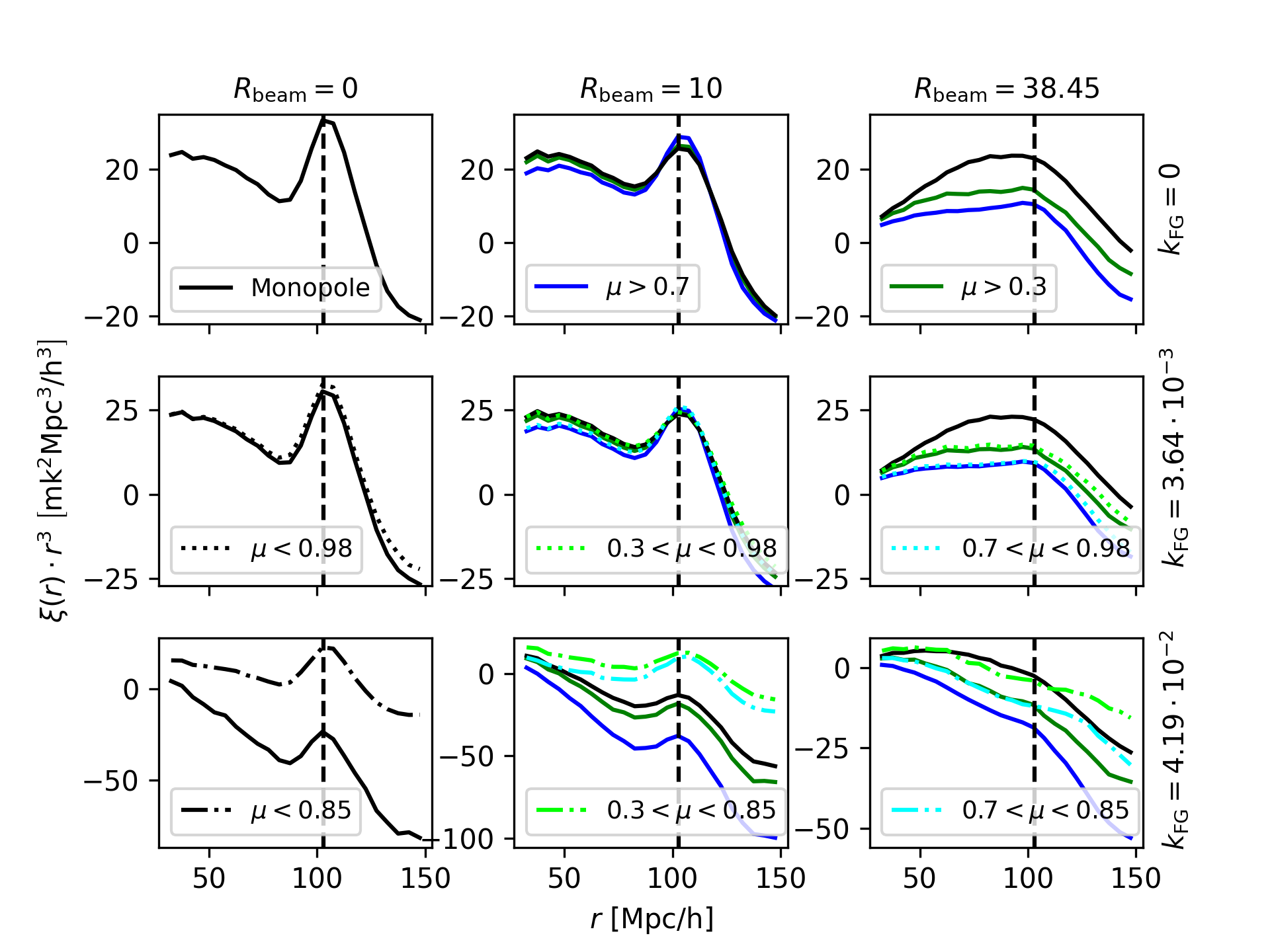}
    \caption{$\mu$-wedge two point correlation functions recovered using all nine combinations of the telescope beam and foreground-removal effects. We consider the monopole ($0<\mu<1$) compared to different combinations of $\mu_{\rm min}=0,\ 0.3,\ 0.7$ and $\mu_{\rm max}=1,\ 0.98,\ 0.85$ adapted to the different cases as discussed in the text. Following the legend, black solid lines represent the monopole, coloured lines always consider a $\mu_{\rm min}>0$ cut, and all dotted and dash-dotted lines consider a $\mu_{\rm max}<1$ cut. The vertical dashed line represents the comoving size of the BAO, $r_{\rm BAO}$, expected for this cosmology.}
    \label{fig:all_wedges}
\end{figure*}

\noindent In this part, we analyse the $\mu$-wedge 2PCF for the combination of both the angular beam and foreground-removal effects in our simulated cosmological signal. Similar to the figures in \secref{sec:HI_2PCF}, we show all combinations of $R_{\rm beam} =\{0,\ 10,\ 38\}\,h^{-1}\,{\rm Mpc}$ (columns) and $k_{\rm FG} = \{ 0, \ 3.64,\  41.9\} \cdot 10^{-3}\,h\, {\rm Mpc}^{-1}$ (rows) in \autoref{fig:all_wedges}. 

For the case without any observational effect (upper-left) we do not consider any cut in $\mu$, resulting simply in the monopole. In the presence of angular smoothing ($R_{\rm beam}>0$), we consider the two $\mu_{\rm min}$ values discussed above (0.3 and 0.7). 
For each foreground-removal case, we consider respective maximum orientations of $\mu_{\rm max}=1,\ 0.98,\ 0.85$, as also discussed above.

For most cases, the BAO signal is clear in all the curves. However, when we inspect the $R_{\rm beam} = 38\,h^{-1}\,{\rm Mpc}$ cases, the BAO signal becomes subtler, especially for the monopoles. Nevertheless, we can recover a feature at the BAO scale when we impose $\mu_{\rm min}=0.3$, which becomes more prominent for $\mu_{\rm min}=0.7$.   
In the presence of foregrounds, whilst $\mu_{\rm max}$ does not help to sharpen the BAO feature, it does help to recover a positive 2PCF, closer to the original purely cosmological signal (top-left).

In general, we find that for different combinations of $R_{\rm beam}$ and $k_{\rm FG}$, the different $\mu_{\rm min}$ and $\mu_{\rm max}$ proposed earlier can help us recover a sharper BAO and partially de-bias the 2-point correlation function. In \secref{sec:nw} 
we focus on the more realistic SKA-like case ($R_{\rm beam} = 38\,h^{-1}\,{\rm Mpc}$, $k_{\rm FG}=\ 3.64\times 10^{-3}\, h\, {\rm Mpc}^{-1}$), and compare the BAO recovery to the purely radial method, which we explain below. For that case, we will also show the error bars of each estimator.

\subsection{Radial 2PCF}
\label{sec:radial}

\paragraph*{Integrated radial 2PCF}
\ \\

\noindent In previous sections we found that the angular smoothing caused by the telescope beam is the dominant effect in erasing the BAO. 
For this reason, \citet{villaescusa} proposed to only consider the radial scales $k_\parallel$ (that work focuses on Fourier space), when looking for BAO in the SKA \hi\ intensity mapping program.
Using the anisotropic 2PCF computed in \secref{sec:HI_2PCF}, we can compute a purely radial 2PCF by averaging over the angular coordinate:
\begin{equation}
\label{eq:radial_integrated}
    \xi (r_\parallel) = \frac{1}{N_{\perp}} \sum_{r_\perp}^{N_\perp}  \xi (r_\perp, r_\parallel) \, ,
\end{equation}
in which $N_{\perp} = r_{\perp, {\rm max}}/l_{\rm grid}$ is the number of angular pixels we average over. In this work, we initially consider this integrated radial 2PCF with
a maximum separation of $r_{\perp, {\rm max}}=150 \,h^{-1}\, {\rm Mpc}$, matched to the limits of \autoref{fig:2PCF2D}. In fact, this new estimator may be seen as a simple sum over the $x$-axis of that figure, keeping the vertical one as the remaining axis.

This estimator can be viewed as complementary to the projected correlation function $w_p(r_\perp)= \int \xi(r_\perp,r_\parallel)\ {\rm d} r_\parallel$, often used in `halo occupation distribution' analysis to integrate out the effect of redshift-space distortions \citep[see e.g.][]{Avila2020}. Hence, it could {\it a priori} be similarly  used here to integrate out the effect of the telescope beam.

\begin{figure*}
    \centering
    \includegraphics[width=1.04\textwidth]{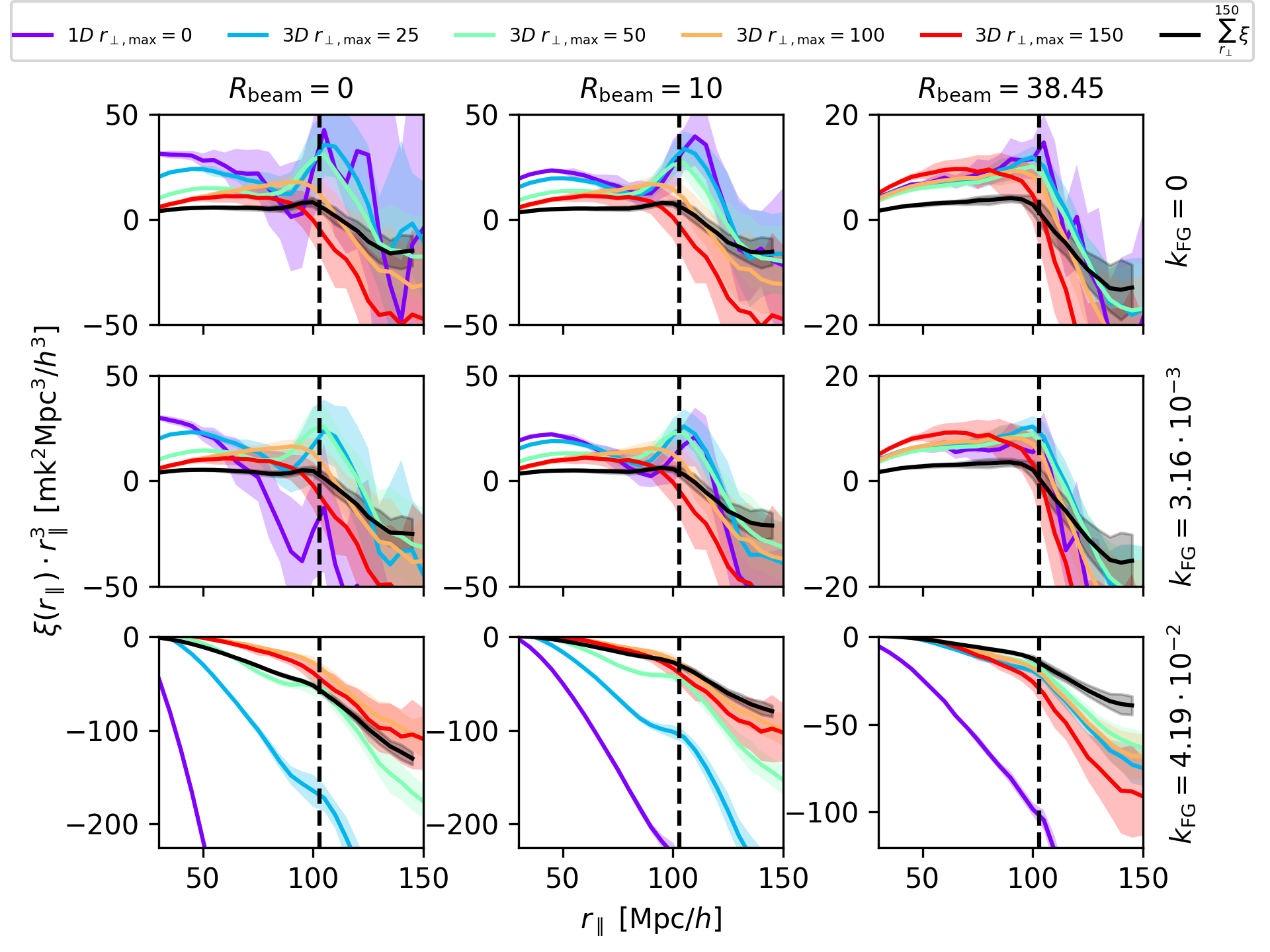}
    \caption{Radial two-point correlation function
    with different definitions used in \secref{sec:radial} for all the combinations of telescope beam sizes and foreground-removal effects. The black lines represent the integrated radial 2PCF, with purple lines the 1D radial 2PCF, and the remaining lines the 3D radial 2PCF with different $r_{\perp,{\rm max}}$, as indicated in the legend. In this figure, we also include a band around each line representing their standard deviation. The vertical dashed line represents the comoving size of the BAO, $r_{\rm BAO}$, expected for this cosmology.
    }
    \label{fig:radial}
\end{figure*}

Following the format of previous sections, we compute this correlation function for the different combinations of $R_{\rm beam}$ and $k_{\rm FG}$, and show them in \autoref{fig:radial} with the $\sum_{r_{\perp}}^{150}\xi$ label. We find that the BAO is visible in most combinations, but not very prominent, even with the lack of observational effects. We find, indeed, this estimator quite insensitive to $R_{\rm beam}$ for the cases $k_{\rm FG}=0$ and $3.64\times 10^{-3}\, h\, {\rm Mpc}^{-1}$, showing its capability to integrate out the effect of the telescope beam, at least partially; for the largest $k_{\rm FG}$, the coupling between radial and angular effects seem to affect the integrated radial 2PCF for different $R_{\rm beam}$ differently. 

We also find that the BAO feature is displaced to lower values due to contributions at $r\simeq r_{\rm BAO}$ with $r_\perp > 0$ and $r_\parallel^2 \simeq r_{\rm BAO}^2 - r_{\perp}^2$. Nevertheless, this does not necessarily imply a bias in the BAO recovery, as long as the theoretical model used to fit the (simulated) data is able to account for it. 

For the discussion of the different radial functions discussed in this subsection, we found it relevant to include error bars to all lines in \autoref{fig:radial} computed as the standard deviation of the 12 combinations given by the four boxes and three orientations (see discussion below \autoref{eq:2PCF_2D}).  
We find that the error bars associated with the integrated radial 2PCF are very small. Hence, even if the feature is less prominent, it is also a promising tool for BAO recovery. 

\paragraph*{1D radial 2PCF} 
\ \\

\noindent The estimator proposed in \citet{villaescusa} uses a 1D radial power spectrum, which is conceptually different to the previous \textit{integrated} method. Following their definition in Fourier space, we define the 1D radial 2PCF in configuration space as:
\begin{equation}
\label{eq:radial1D}
    \xi_{\rm 1D}(\lvert \vec r_{\parallel} \lvert) = \left\langle \Delta T (\vec X_\perp, \vec X_\parallel) \cdot \Delta T (\vec X_\perp, \vec X_\parallel + \vec r_\parallel)  \right\rangle_{\forall \vec X_\perp} \, ,
\end{equation}
where the average is performed over all lines of sight $\vec X_\perp$, where for each line of sight, all pairs of pixels separated by a distance $ r_{\parallel} =\lvert \vec{r}_{\parallel} \lvert$ contribute. In this case, each line of sight is given by the pixelisation described in \secref{sec:intensity_maps} with $l_{\rm grid}=5\, h^{-1}\, {\rm Mpc}$ and we use again $\Delta r_{\parallel} = 5\, h^{-1}\, {\rm Mpc}$, which simplifies the computations. 

We also represent this estimator in \autoref{fig:radial} (1D) for all combinations of observational effects. We find that this estimator shows a very prominent BAO feature, but at the same time it can be very noisy, especially when no angular smoothing is applied.

\paragraph*{3D radial 2PCF} 
\ \\

\noindent In search for something less noisy than the 1D radial correlation and with a more prominent BAO feature than the integrated radial 2PCF, we define the following estimator:
\begin{equation}
\label{eq:radial_max}
    \xi_{r_{{\perp},{\rm max}}}(r_{\parallel}) = \left\langle \Delta T ( \vec X_\perp , \vec X_\parallel)\cdot  \Delta T ( \vec X_\perp + \vec r_\perp , \vec X_\parallel + \vec r_\parallel ) \right\rangle_{ r_\perp<r_{\perp,{\rm max}}} \, ,
\end{equation}
where we average over all pairs of pixels that are separated by a given $r_\parallel$ with a maximum angular separation ($r_\perp  \equiv \lvert \vec{r}_\perp \lvert < r_{\perp, \text{max}}$). 

This estimator is related to the integrated radial function, but it is not exactly the same. If we used the same $r_{\perp,{\rm max}}$ for both estimators and introduced a weight in \autoref{eq:radial_integrated}  proportional to the number of pairs of $\Delta T$ pixels that enter in each $r_{\perp}$ bin, we would recover the  $\xi_{r_{{\perp},{\rm max}}}(r_{\parallel})$ estimator. However, the latter may be computed directly from the $\Delta T$ grid very efficiently, and the same code can be used to obtain the 1D estimator by using $r_{\perp,{\rm max}}=0$.

Additionally, these related estimators can give us different information. Following the analogy between the integrated radial 2PCF and the projected correlation ($w_p$), the $\xi_{r_{{\perp},{\rm max}}}(r_{\parallel})$ estimator (\autoref{eq:radial_max}) could be interpreted as the reciprocal to the angular correlation function $w(\theta)$ of a tomographic $z$-bin in a galaxy sample. 

In \autoref{fig:radial} we plot the $\xi_{r_{\perp,{\rm max}}}(r_{\parallel})$ estimator for $r_{{\perp},{\rm max}}=\{25,50,100,150\} h^{-1}\,{\rm Mpc}$ (also for $r_{{\perp},{\rm max}}=0$, but we refer to this one as the 1D correlation).
We see that by increasing $r_{{\perp},{\rm max}}$ we reduce the noise in the correlation with respect to the 1D case, but at the same time, the BAO feature becomes less prominent, nearly disappearing for some of the configurations. Following the previous analogy, this is equivalent to increasing the width of a redshift bin when studying the BAO in the angular correlation function \citep[$w(\theta)$, e.g.][]{angularBAO}. 

\subsection{BAO in a SKA-like configuration}
\label{sec:nw}
In the previous subsections we studied a variety of different methods aimed at isolating the BAO for different combinations of telescope beam size and foreground-cleaning effects. Here we compare the most promising estimators for the case considered to be the most representative of a future SKA intensity mapping experiment: $R_{\rm beam}=38\,h^{-1}\,{\rm Mpc}$, $k_{\rm FG}=3.16\times 10^{-3}\, h\,{\rm Mpc}^{-1}$.

\begin{figure}
    \centering
    \includegraphics[width=0.52\textwidth]{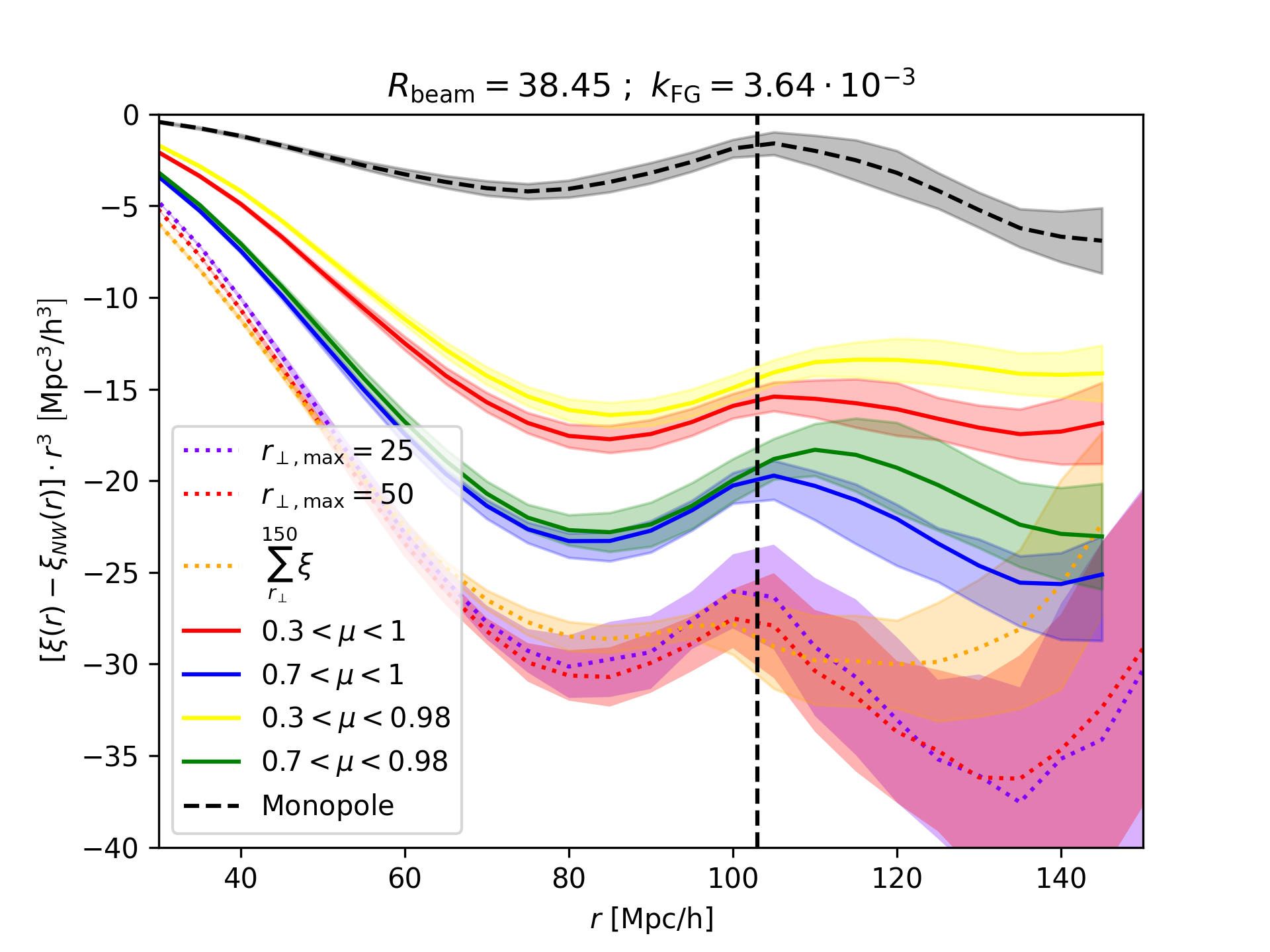}
    \caption{Summary of different 2PCFs proposed for an SKA-like case ($R_{\rm beam}=38\,h^{-1}\, {\rm Mpc}$ and $k_{\rm FG}=3.64 \times 10^{-3} h\,{\rm Mpc}^{-1}$). In this plot, we subtracted the non-wiggle model $\xi_{\rm nw}$ defined in \autoref{eq:nw} from the different 2PCF estimators in order to highlight the BAO signal. We include the monopole (dashed), four different $\mu$-wedges (solid), the integrated radial 2PCF and the 3D radial 2PCF with different $r_{\perp, {\rm max}}$ (all radial functions in dotted lines).
    }
    \label{fig:nonBAO}
\end{figure}

To better highlight the BAO feature, we subtract from our simulation's 2PCF a model without BAO (referred to as a non-wiggle, "{\it nw}" model) derived from the \citet{Eisenstein&Hu} power spectrum obtained with \textsc{camb}%
\footnote{\url{https://camb.info/sources/}; \citet{camb}.}.
We include a free linear bias parameter $b_\hi$, an angular beam following \citet{cunnington20}, and apply the usual Hankel transform to obtain the model 2PCF:
\begin{multline}
    \label{eq:nw}
    \xi_{\rm nw} (r) = \int_0^\infty \frac{k^2}{2\pi^2}\, j_0(kr)\, T_b^2\, b_{\hi}^2\, P_{\rm EH}(k)\,  {\rm d}k \\
    \times \int_0^1 \exp\left[\frac{-k^2\, R_{\rm beam}^2\, \left(1-\mu^2\right)}{2}\right]\, {\rm d} \mu \, .
\end{multline}
Note that this computation, for the sake of simplicity, corresponds to the monopole and does not include the effect of foreground cleaning or any of the geometrical tricks described in this paper. We find that this template works well to highlight the BAO in \autoref{fig:nonBAO} and we leave a more detailed and precise model fit  for future work. We also note that this type of non-wiggle template is typically accompanied by several power-law terms that remove non-BAO-like residuals \citep[e.g.][]{DESY1BAO}. 

In \autoref{fig:nonBAO} we find that the BAO feature is visible for all the lines, even for the monopole (black), which is the simplest estimator for the 2PCF. 
We see that the integrated radial correlation function ($\sum_{r_\perp}^{150} \xi_\parallel$) smooths out the BAO too much, whereas the radial functions with a mild $r_{\perp}$ allowance ($<50$ or $<25$) work much better in getting a sharp BAO. These functions, however, have a large noise level: the smaller the $r_{{\perp}, {\rm max}}$, the sharper but noisier the BAO feature is. On the other end, the $\mu$-wedge 2PCF have tighter error-bars but with a less prominent peak. Out of the latter, the most promising wedge seems to be the $0.7<\mu<1$ curve (blue). A more quantitative study will need to be done in the future to determine how the trade-off between sharpness of the BAO, uncertainty on the 2PCF, and possible biases, is reflected in the final determination of the BAO distance $r_{\rm BAO}$. 
A simple visual inspection seems to indicate a bias of the $\sum_{r_\perp}^{150} \xi_\parallel$ estimator (orange) toward lower values of the BAO peak and of the $\mu_{\rm max}=0.98$ wedges (yellow and green) toward larger values for the BAO feature. However, this could be easily due to the different integration and projection effects, or the choice of representation for the $y$-axis (with $\times\ r^3$). Once a more detailed and accurate model is constructed for these estimators, it should capture all these effects. Hence, when fitting for BAO using a template based  in those models, the possible biases in the  $r_{\rm BAO}$ should go away.

To sum up, we find that all the curves proposed here have the potential to be a good tool to detect BAO with SKA. We emphasise the potential of the $\mu$-wedges (for the first time studied in the context of IM) and on the 3D radial 2PCF introduced here. However, we leave for future work a more quantitative comparison of the different methods once a theoretical model is constructed and validated for those estimators, together with a covariance and fitting pipeline.
Additionally, the choice of the best method may depend on the $R_{\rm beam}$ and $k_{\rm FG}$ parameters. Whereas here we focused on a SKA-like experiment, it would be interesting to study other cases that may be more suited for other experiments.

\section{Summary and Conclusions}
\label{sec:conclusions}

In this work we analysed the \hi\ intensity mapping clustering by measuring different types of two-point correlation functions from simulations. We focused on the anisotropy induced by the telescope beam and the foreground removal techniques, and on studying different techniques to recover the BAO.

For that, we used the $z=1.321$ snapshot of the UNIT simulations, a suite of four $1\,h^{-1}\,{\rm Gpc}$ boxes that uses the Fixed \& Pair technique in order to enhance the effective volume to $\sim\!150\,h^{-3}\,{\rm Gpc}^3$ \citep{Chuang}. Additionally, these simulations have had the SAGE semi-analytic model of galaxy formation run on them \citep{sage,Knebe21}, giving a galaxy catalogue that includes cold gas masses, from which we derive \hi\ masses using prescriptions based on \citet{Baugh04} (\autoref{eq:Rmol04}, default) and \citet{blitz} (\autoref{eq:Blitz}). From these simulations we studied the \hi--halo mass relation, finding a similar curve to other ones proposed in the literature. With those two prescriptions we made a prediction for the total \hi\ abundance\,---\,$\Omega_\hi (z=1.321)=3.8\times 10^{-4}$\ \& $\Omega_\hi (z=1.321)=2.7\times 10^{-4}$\ , respectively ---\,and discussed how the simulation mass resolution or the uncertainty in the \hi--halo relation could affect this value. All the values found are within 2-$\sigma$ of the observational measurements by  \citet{Rao2006}. 

We then described in \secref{sec:HI_2PCF} the methodology used to create \hi\ intensity maps and the methods to simulate (in an approximate but fast way) the addition of observational effects produced by the telescope beam and foreground cleaning. These maps were then analysed by means of the anisotropic 2-point correlation function, $\xi(r_\perp, r_\parallel)$, for different combinations of observational effects ($R= \{0, 10, 38\} h^{-1}\, {\rm Mpc}\  \otimes\ k_{\rm FG} = \{ 0,\ 3.64,\ 41.9\}  \times 10^{-3} h\,{\rm Mpc}^{-1}$). We found these 2PCFs crucial in understanding the impact of the observational effects, as they affect the distances parallel ($r_\parallel$) and perpendicular ($r_\perp$) to the line of sight very differently. We find that, whereas the effect of foregrounds tend to be very localised close to the line of sight, the effect of the beam can affect all scales, specially for the larger beam size considered (expected for SKA at $z=1.321$). Nevertheless, the BAO signature is still apparent in all those cases. This is the first study that looks at these effects with the $\xi(r_\perp, r_\parallel)$ 2PCF computed on simulations. In parallel, an independent team studied the same function [$\xi(r_\perp, r_\parallel)$] from the theoretical modelling side. Their results were recently published \citep{Knennedy&Bull} and simple visual comparison yields an excellent agreement. More direct and quantitative comparison will be pursued in future studies. 

Based on the apparent BAO signal in the anisotropic 2PCF, in \secref{sec:BAO}, we explored different summary 2PCFs able to isolate the BAO signal. We studied a $\mu$-wedge 2PCF bounded by a $\mu_{\rm min}$ and $\mu_{\rm max}$ that can be tuned to avoid the orientations heavily affected by the observational effects. This 2PCF is compared against the reference monopole with $\mu_{\rm min}=0$ and $\mu_{\rm max}=1$.  We find that foregrounds cause a general lowering of the 2PCF that can be mostly corrected by setting $\mu_{\rm max}=0.98$ and $0.85$ for $k_{\rm FG} =  3.64$ and $41.9\times 10^{-3}\, h\,{\rm Mpc}^{-1}$, respectively. However, the BAO peak seems unaffected by foregrounds. For the beam effect, we find that the BAO feature progressively sharpens as we increase $\mu_{\rm min}$. For the larger beam size ($R_{\rm beam}=38\,h^{-1}\, {\rm Mpc}$), we may need to consider a cut as drastic as $\mu_{\rm min}=0.7$.

We also studied a family of radial correlation functions. On one end, we studied the integrated radial function $\sum_{r_{\perp}}^{150} \xi$, in which the BAO is mostly removed due to the smoothing performed over such a wide range of scales. On the other end, we measured the 1D radial correlation function motivated by the 1D radial power spectrum proposed in \citet{villaescusa}, in which a very sharp but noisy BAO feature is found. We then find a middle ground on a 3D radial correlation function with a maximum angular separation given by $r_{\perp, {\rm max}}=25$ or $50\,h^{-1}\,{\rm Mpc}$. 

Finally, we compare all the proposed summary 2PCF in \autoref{fig:nonBAO} for the case most representative of an SKA1-MID \hi\ intensity-mapping survey with $R_{\rm beam}=38\,h^{-1}\, {\rm Mpc}$ and $k_{\rm FG} = 3.64  \times 10^{-3}\, h\,{\rm Mpc}^{-1}$. We find that all the proposed 2PCFs show a clear BAO feature with different levels of prominence and uncertainty. 

We leave for future work the theoretical modelling and statistical analysis required to determine which of those estimators can provide the most accurate and precise measurements of BAO and cosmological parameters. \citet{Knennedy&Bull} performed a statistical analysis for redshift-space \hi\ IM multipoles based on theoretical models for a MeerKLASS-like experiment, which is very complementary to this work.
Whereas here we studied the \hi\ clustering in real space in order to highlight the effect of the instrumental effects on the anisotropy,  implementing redshift-space distortions would be straightforward.
Combined analyses of simulations and models should follow up on the aforementioned paper and this work.

In the next few years, a number of experiments will measure the clustering of \hi\ via intensity-mapping programs. We remark that some of them will have milder telescope beams and, hence, the detection of BAO will simplify significantly. SKA operations are a decade away and, in the meantime, we expect to refine many of the techniques to improve our analysis techniques. The detection of BAO with \hi\ IM will be a milestone in the maturity of this field as it was in the field of galaxy surveys two decades ago. Nevertheless, this is only a step towards unveiling the real potential of the \hi\ IM, which will be able to probe unexplored scales close to the size of the observable Universe. 

\section*{Acknowledgements}

We thank Alkistis Pourtsidou for enlightening discussions, specially at early stages, and Violeta Gonz\'{a}lez-P\'{e}rez and Phil Bull for very useful feedback. We also thank Kazim Çamlibel for feedback on the references.
SA is supported by the MICUES project, funded by the EU H2020 Marie Skłodowska-Curie Actions grant agreement no. 713366 (InterTalentum UAM). 
BVG is supported by the Atracci\'{o}n de Talento Contract no. 2019-T1/TIC-12702 granted by the Comunidad de Madrid in Spain. He also has been funded by a JAE-INTRO 2019 studentship no. JAEINT19\_EX\_1003, provided by CSIC and extended by IFT (UAM-CSIC).
SC is supported by STFC grant ST/S000437/1. 
ARHS acknowledges receipt of the Jim Buckee Fellowship at ICRAR-UWA.
GY and AK are supported by the Ministerio de Ciencia, Innovaci\'{o}n y Universidades (MICIU/FEDER) under research grant PGC2018-094975-C21. AK further thanks Lambchop for `Is a Woman'.

The UNIT simulations have been run in the MareNostrum supercomputer at the Barcelona Supercomputer Center thanks to computing time awarded by the PRACE consortium under project grant 2016163937.

{\it Author contributions}: SA \& BVG co-led the project and paper. SC made key contributions to the project and paper. AS, GY, AK and CHC contributed to the paper and led the construction of the simulations used.

\section*{Data Availability}

The simulations used in this paper are publicly available at \url{http://www.unitsims.org}, where users can find both dark matter halo and SAGE galaxy catalogues.  The SAGE code-base itself is publicly available at \url{https://github.com/darrencroton/sage}. Other numerical results may be shared under reasonable request to the corresponding authors.



\bibliographystyle{mnras}
\bibliography{BAOwithSKA} 




\appendix

\section{The $M_\hi(M_h )$ relation: central-satellite split and new analytical fits}
\label{sec:fit}

\begin{figure*}
    \centering
    \includegraphics[width=0.52\textwidth]{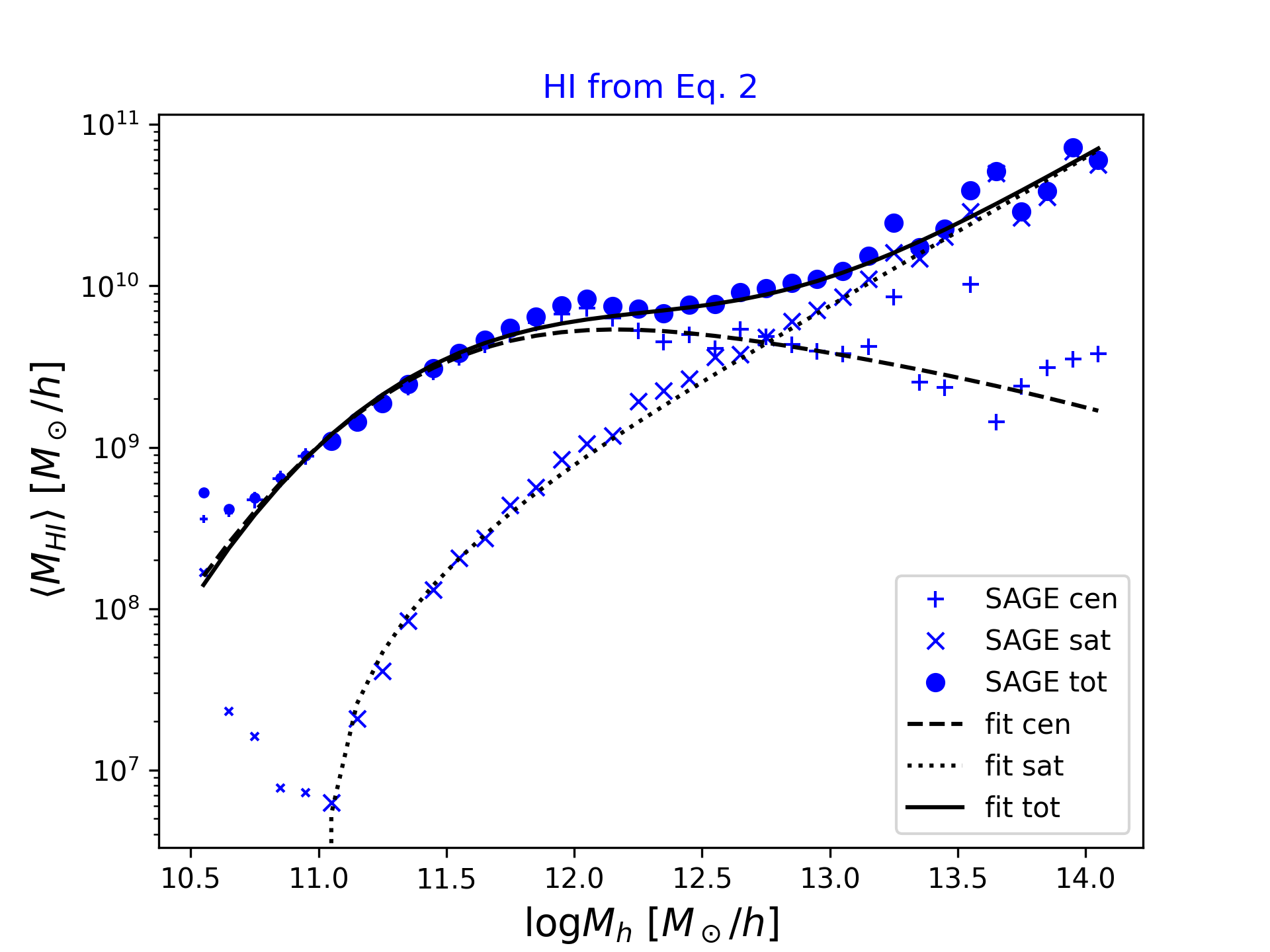}\includegraphics[width=0.52\textwidth]{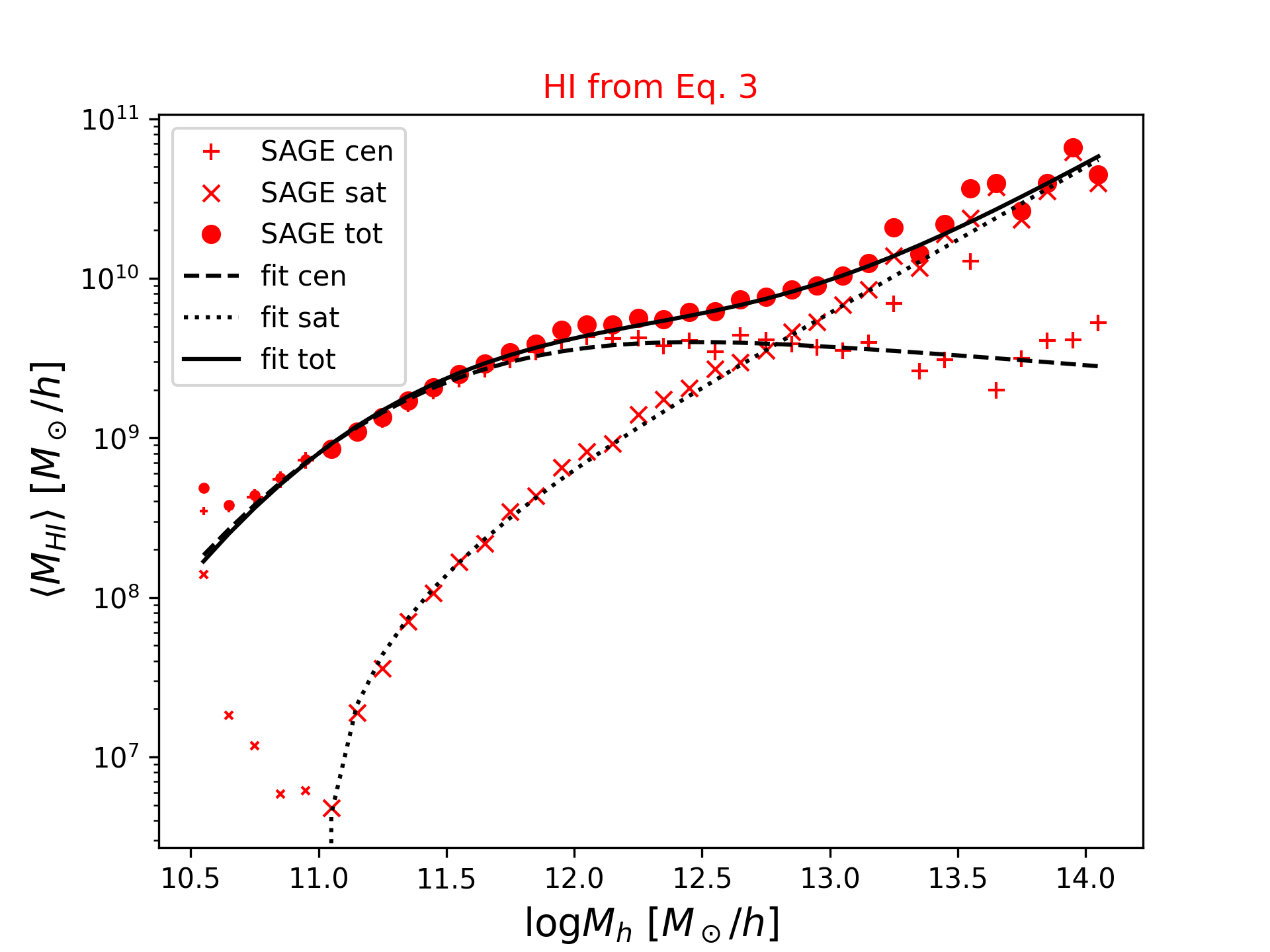}
    \caption{\hi\ Halo Occupation Distributions from the SAGE-\unit\ catalogues and analytical fits to them. We represent for each halo mass interval (whose center is the $x$-axis) the mean total (`tot') \hi\ mass ($y$-axis) as well as the separated contribution from central galaxies (`cen') and from satellite galaxies (`sat'). Whereas the different types of points represent the measurements from the SAGE, the lines represent analytical fits with Eqs. \ref{eq:central} \& \ref{eq:sat} and the best fit parameters reported in \tabref{tab:hod}. The first two points for the satellites and the first five points for the centrals are not included in the fits, since they seem to be affected by resolution effects. These points are represented with a smaller size. {\bf Left:} \hi\ masses derived from \autoref{eq:Rmol04}. {\bf Right:} \hi\ masses derived from \autoref{eq:Blitz}. 
    }
    \label{fig:hod}
\end{figure*}


In \autoref{fig:heatmap}, we presented a 2D heat-map histogram of the halo masses $M_h$ and the total \hi\ mass $M_\hi$ in each halo as found in the SAGE-\unit\ catalogues. We also represented in black lines the mean \hi\ mass in each halo mass bin [$M_\hi(M_h)$]. 

In this Appendix, in \autoref{fig:hod}, we split the \hi\ contribution coming from central and satellite galaxies, as it is typically done in Halo Occupation Distribution (HOD) studies. We perform this analysis for both the $R_{\rm mol}=0.4$ (\autoref{eq:Rmol04}, left) and the \citet{blitz} (\autoref{eq:Blitz}, right) \hi\ prescriptions.  

In order to make this HI HOD information more useful for the community, we have fitted analytical curves to the SAGE-\unit\ results. In these fits we have discarded the points that seem affected by resolution effects: the first 2 points for the centrals and the first 5 points for the satellites.  

For central galaxies we use

\begin{equation}
\label{eq:central}
    \langle M_{\hi,\, {\rm cen}}(M_h) \rangle  = \frac{1}{2}  \Bigg[ 1+ {\rm erf}\Bigg( \frac{{\rm log}M_h - {\rm log} M_0}{\sigma}\Bigg)\Bigg]\cdot M_{\hi}\  e^{-\Big(\frac{M_h}{M_0}\Big)^\gamma} \, ,
\end{equation}
where the first part (up to "$\cdot$") corresponds to the standard 5-parameter HOD model \citep{Zheng05} and we introduce a second part to better fit $ \langle M_{\hi,\, {\rm cen}}(M_h) \rangle$. 
This new part contains a normalisation factor ($M_\hi$) and an exponential damping that suppresses the $\hi$ occupation at large halo masses (see \autoref{sec:HIprescription} for a discussion on the physical interpretation of this suppression). The parameters $M_0$, $\sigma$, $M_\hi$ and $\gamma$ are fitted to the results from SAGE-\unit\ and their best fit values are shown in \tabref{tab:hod}.

We do something similar to fit the HI HOD satellite component. We use the following function: 

\begin{equation}
\label{eq:sat}
    \langle M_{\hi,\, {\rm sat}} (M_h) \rangle = \Bigg( \frac{M_h-M_0}{M_1}\Bigg)^\alpha \cdot \frac{1}{2}\Bigg[ 1+ {\rm erf}\Bigg( \frac{{\rm log}M_h - {\rm log} M_0}{\sigma}\Bigg)\Bigg]\,  M_\odot/h \ .
\end{equation}

Again, we start by including the standard power-law (the part before the "$\cdot$" corresponds to \citealt{Zheng05}), whereas in the second part we introduce a smoothing factor that, in this case, helps fitting the low-mass end. The best fit value for the $M_0$, $M_1$, $\alpha$ and $\sigma$ parameters are also shown in \tabref{tab:hod}. Note that, whereas some parameters are called the same in \autoref{eq:central}
and \autoref{eq:sat}, they are treated independently to fit separately the central and satellite HOD. 

One feature we find is that, when using the $R_{\rm mol}=0.4 $ (\autoref{eq:Rmol04}) prescription, we obtain a more pronounced \hi\ suppression for centrals at large halo masses.
Finally, we note that there is a small under-fitting around the peak of the central HI HOD. We tried to correct this with different functional forms of $\langle M_{\hi,\, {\rm cen}} (M_h) \rangle$ with the same number of parameters, but the under-fitting remained. We believe that this could be corrected by including statistical error bars accounting for Poisson noise in the bins, significant at the high mass end. Alternatively, a Gaussian curve or other peaked curved could be added to account for the current residuals. However, this would imply increasing the already large number of parameters. Either of those solutions is beyond the scope of this study and we consider these fits sufficiently accurate, given the uncertainties in the $M_\hi$-$M_h$ relation discussed in \autoref{sec:HIprescription}. 

\begin{table}
    \centering
    \begin{tabular}{l|c|c|c|c}
    \hline
         Data & ${\rm log } M_0$ &  ${\rm log } M_1$ or ${\rm log } M_\hi$ & $\sigma$ & $\alpha$ or $\gamma$  \\
           \hline
          \hline
          Centrals Eq. 2 & 11.60 &. 10.32 & 0.680 & 0.1639\\
         Satellites Eq. 2 & 11.02 & 2.20 & 0.939 & 0.915 \\
         Centrals Eq. 3 & 11.60 & 10.14 & 0.789 & 0.0820\\
         Satellites Eq. 3 & 11.02 & 2.24 & 0.984 & 0.910 \\

           \hline
    \end{tabular}
    \caption{Parameters of Eqs. \ref{eq:central} \& \ref{eq:sat} that best fit the $\langle M_\hi (M_h)\rangle$. relation for the central and satellite components and for the two \hi\ prescriptions considered in our catalogues. The best fit is computed by minimising the quadratic sum of the relative differences between the analytical curves (`th') and the SAGE-\unit\ results (`SAGE') in the $M_h$ range considered (see text and \autoref{fig:hod}): $\sum \big(M_\hi^{\rm th} - M_\hi^{\rm SAGE}\big)^2/\big(M_\hi^{\rm SAGE}\big)^2$ .}
    \label{tab:hod}
\end{table}

\section{The $M_\hi(M_h )$ relation: fits used from the literature}
\label{sec:analytical}

In this Appendix we collect the analytical expressions of the different halo mass - \hi\ mass relations $M_\hi(M_h)$ taken from the literature for our comparison in \autoref{fig:heatmap} \& \autoref{fig:omegas}. 

The first one has the following expression \citep[][]{moster,padmanabhan}:

\begin{equation}
    M_{\hi,1}=2 N_1 M_h \left[\left(\frac{M_h}{M_1}\right)^{-b_1}+\left(\frac{M_h}{M_1}\right)^{y_1}\right]^{-1}
\end{equation}

This shape was first introduced by \citet{moster} in order to relate stellar mass to halo mass and was later reused by \citet[][]{padmanabhan} in order to relate \hi\ to halo mass. It has four redshift dependent parameters: the normalization $N_1$, the characteristic mass $M_1$ and two slopes  $b_1$ and $y_1$, that control the sharpness of the decay (see discussion of the physics behind this in \autoref{sec:HIprescription}). The reported redshift dependence of the parameters is further parameterized as \citep{padmanabhan}:

\begin{equation}
     \begin{split}
    {\rm log}M_1 = {\rm log} M_{10}+\frac{z}{z+1} M_{11}\, , \\
    N_1 = N_{10} + \frac{z}{z+1} N_{11}\, , \\
    b_1 = b_{10}+\frac{z}{z+1}b_{11}\hspace{0.5cm} {\rm \&} \\
    y_1 = y_{10} + \frac{z}{z+1}y_{11}\, .
    \end{split}
\end{equation}

In \citet{padmanabhan}, they fitted those parameters to low and high redshift measurements, reporting  $M_{10} = \left(4.58 \pm 0.19 \right)\cdot 10^{11} M_\odot$, $N_{10} = \left(9.89 \pm 4.89 \right)\cdot 10^{-3} $, $b_{10} = 0.90$, $M_{11} = 1.56^{+0.53}_{-2.70}$, $N_{11} = 0.009^{+0.06}_{-0.001}$, $b_{11} = -1.08^{+1.52}_{-0.08} $ and $y_{11} = 4.07^{+0.39}_{-2.49}$.

The second \hi\ mass assignment we consider was proposed by \citet{bagla}:

\begin{equation}
\label{eqn:bagla}
M_{\hi,2} = \left\lbrace
\begin{array}{ll}
 \frac{f_2 M_h}{1+\left(\frac{M_h}{M_{\rm max}}\right)^2} &  \textup{if } M>M_{\rm min} \\
  0 & \textup{if } M<M_{\rm min}
\end{array}
\right.
\end{equation}

with $M_{\rm{min}}= 1.92 \cdot 10^9 M_\odot/h$, $M_{\rm{max}} = 5.68 \cdot 10^{11} M_\odot/h$ and $f_2 = 0.0159$.

The third parametrisation of HI mass we consider was introduced by \citet{baugh}:

\begin{equation}
    M_{\hi,3}=M_h \left[a_1\left(\frac{M_h}{10^{10}}\right)^\beta e^{-\left(\frac{M_h}{M_{\rm break}}\right)^\alpha}+a_2\right] \, ,
\end{equation}
in which the values of the parameters are (we use the values fitted at $z=1$): $a_1 = 7 \cdot 10^{-3}$, $a_2 = 1.5\cdot 10^{-4}$, ${\rm log}\left(M_{\rm break}\right) = 11.5$,$\alpha = 1.5$ and $\beta = 0.2$.

Finally, we also consider an extension of \citet{baugh}, propossed by \citet{spinelli} in order to include a cut-off at low masses. The expression is as follows:

\begin{equation}
    M_{\hi,4}=M_h \left[a_1\left(\frac{M_h}{10^{10}}\right)^\beta e^{-\left(\frac{M_h}{M_{\rm break}}\right)^\alpha}+a_2\right]e^{-\left(\frac{M_{\rm min}}{M_h}\right)^\gamma}
\end{equation}

in which the values of the new parameters are $\gamma = 0.5$ and ${\rm log} M_{\rm min} = -1.3$. The other redshift-dependent parameters were re-calibrated: $a_1 = 3.8 \cdot 10^{-3}$, $a_2 = 1.6 \cdot 10^{-3}$, $\alpha = 0.24$, $\beta = 1.70$, ${\rm log} M_{\rm break} = 8.30$ \citep{spinelli}.


\bsp	
\label{lastpage}
\end{document}